\documentclass{aa}
\usepackage{txfonts}
\usepackage{natbib}
\usepackage{graphicx}

\bibpunct[, ]{(}{)}{;}{a}{}{,}

\def\logg{\log g}
\def\teff{T_{\rm eff}}

\def\kms{km/s}

\def\synthmag{{\sc Synthmag}}
\def\synmast{{\sc Synmast}}

\def\loggf{$\log(gf)$}

\def\vsini{$\upsilon\sin i$}

\def\hunda{Hund's case~(a)}
\def\hundb{Hund's case~(b)}
\def\br{B_{\rm r}}
\def\bm{B_{\rm m}}
\def\ba{B_{\rm a}}
\def\b{|\vec{\mathrm{B}}|}
\def\bv{\vec{\mathrm{B}}}
\def\btimesf{(|\vec{\mathrm{B}}|f)}

\begin{document}

\title{Modelling the molecular Zeeman-effect in M-dwarfs: methods and first results.}

\author{D. Shulyak\inst{1}, A. Reiners\inst{1}, S. Wende\inst{1}, O. Kochukhov\inst{2}, N. Piskunov\inst{2} \and A. Seifahrt\inst{1,3}}
\offprints{D. Shulyak, \\
\email{denis.shulyak@gmail.com}}
\institute{%
Institute of Astrophysics, Georg-August-University, Friedrich-Hund-Platz 1, D-37077 G\"ottingen, Germany \and
Department of Physics and Astronomy, Uppsala University, Box 516, 751 20, Uppsala, Sweden \and
Department of Physics, University of California, One Shields Avenue, Davis, CA 95616, USA
}

\date{Received / Accepted}

\abstract
{}
{We present first quantitative results of the surface magnetic field measurements in selected M-dwarfs 
based on detailed spectra synthesis conducted simultaneously in atomic and molecular lines of the FeH Wing-Ford $F^4\,\Delta-X^4\,\Delta$ transitions.}
{A modified version of the Molecular Zeeman Library (MZL) was used to compute Land\'e g-factors for FeH lines in different Hund's cases. 
Magnetic spectra synthesis was performed with the \synmast\ code.}
{We show that the implementation of different Hund's case for FeH states depending on their quantum
numbers allows us to achieve a good fit to the majority of lines in a sunspot spectrum in an automatic regime.
Strong magnetic fields are confirmed via the modelling of atomic and FeH lines for three M-dwarfs
\object{YZ~CMi}, \object{EV~Lac}, and \object{AD~Leo}, but their mean intensities are found to be systematically lower
than previously reported.
A much weaker field ($1.7-2$~kG against $2.7$~kG) is required to fit FeH lines in the spectra of \object{GJ~1224}.}
{Our method allows us to measure average magnetic fields in very low-mass stars from polarized radiative transfer. 
The obtained results indicate that the fields reported in earlier works were probably overestimated by about $15-30$\%.
Higher quality observations are needed for more definite results.}


\authorrunning{Shulyak et al.}

\maketitle

\section{Introduction}
\label{sec:intro}
Magnetic fields in non-degenerate stars are found all across the
Hertzsprung-Russell diagram, from hot high-luminous stars down to cool
and ultra-cool dwarf \citep[see, for example, the review by][and
references therein]{donati-landstreet2009}.  These fields spawn a wide
range of intensity and geometry, thus providing strong experimental
ground for the theories of stellar magnetism.

Contrary to the organized, large-scale magnetic fields of early- and
intermediate-type stars \citep[e.g.][]{landstreet1992,landstreet2001}
that are probably of fossil origin, the magnetic fields of low-mass
stars are believed to have a dynamo-generated nature.  These stars
often show activity in their atmospheres similar to the Sun
\citep{berdyugina2005}, however, objects later than M$3.5$ are
believed to become fully convective, therefore different dynamo mechanisms
need to be involved to explain their fields. The characterizations of
the magnetic fields in these objects are of great importance for the
general understanding of its generation and evolution.

Direct measurements of the magnetic fields in cool stars relie on
Zeeman broadening of spectral lines \citep[e.g.][]{saar2001}. Strong
fields up to $\approx4$~kG were then reported for some M-dwarfs based
on the relative analysis of magnetically sensitive atomic lines
\citep{jk-valenti1996,jk-valenti2000}. For dwarfs cooler than mid~-~M, 
atomic lines
decay rapidly and are lost in the forest of molecular features. As a
result, the search for alternatives ended up with molecular lines of
FeH Wing-Ford $F^4\,\Delta-X^4\,\Delta$ transitions around
$0.99~\mu$m \citep{valenti2001,r-and-b-2006}.  Some of these lines do
show strong magnetic sensitivity, as seen, for instance, in the
sunspot spectra \citep{wallace1998}.

The modelling of the Zeeman effect in FeH lines though faces a
great difficulty: most lines are formed in the intermediate Hund's
case, the theoretical description of which is based on certain approximations.
Among recent improvements one must mention the work of \citet{berdyugina2002}, 
who extended the formulae initially provided by \citet{schadee1978}, but still within the
Zeeman regime; and the further theoretical development by \citet{ramos2006}, who
presented a very general approach of computing the effect of the magnetic field
on electronic states with arbitrary multiplicity in incomplete Paschen-Back regime.
The main problem is connected with the Born-Oppenheimer approximation,
which was used in theoretical descriptions of level splitting, and
which assumes a clear separation between the electronic and the nuclear motion in terms of energies.
This approximation fails for FeH because the energy separation between the electronic states is
of the order of or smaller than the energy separation between individual vibrational levels.
We refer the interested reader to the aforementioned papers for more details.
A promising solution was then suggested by
\citet{r-and-b-2006,r-and-b-2007}, who estimated the magnetic fields
(or, more precisely, product $\btimesf$, where $f$ is a filling
factor) in a number of M-dwarfs by simple linear interpolation between
the spectral features of two reference stars with known magnetic
fields.  They confirmed the presence of rather strong fields of the
order of $2-4$~kG in a number of M-dwarfs, but the error bars of such
an analysis stays high ($\approx1$~kG).

Later, \citet{afram2008} made use of a semi-empirical approach to
estimate the Land\'e g-factor of FeH lines in the sunspot spectra.
The authors succeeded to obtained a very good agreement with
observations for selected FeH lines and presented best-fitted
polynomial g-factors of upper and lower levels of corresponding
transitions. A little earlier, \citet{harrison-brown2008} presented
the empirical g-factors for a number of FeH lines originating from
levels with rotational quantum numbers $\Omega=7/2, 5/2, 3/2$, but
limited to low magnetic $J$-numbers.
The authors also presented a way in which the effective Hamiltonian
approach can still be used by modifying electronic spin $g_{\rm S}$ 
and orbital magnetic $g_{\rm L}$ factors from their theoretical values.
These studies clearly showed
the main problems of the modern theory of the Zeeman effect in
intermediate Hund's case and pointed in the direction of combining
empirical and theoretical approaches.

The empirical analysis of FeH lines presented in
\citet{r-and-b-2006,r-and-b-2007} is the only estimate of the magnetic
field using molecular lines in M-dwarfs available so far.  Yet the procedure
employed in these studies is not physically justified and thus
requires more quantitative investigation. Further analyses of FeH
lines also employed the same method to measure M-dwarf magnetic fields
\citep[e.g.,][]{r-etal-2009, r-and-b-2009, r-and-b-2010}. It is
important to realize that so far all field measurements in FeH lines
are anchored in the measurement of the field strength of EV~Lac,
$\btimesf = 3.9$\,kG, which was carried out in a single atomic Fe line
by \cite{jk-valenti2000}. Here, we attempt to provide an independent
measurement from a number of magnetically very sensitive molecular FeH
lines, which are modelled based on the formalism described in \citet{berdyugina2002}. 
Our main goal is to combine the information from available atomic and molecular
diagnostics employing the direct spectrum synthesis. This would
provide a more consistent information about the magnetic fields in
these interesting objects and thus provide a step forward towards our
understanding of their magnetic properties.

\section{Observations}

In the present analysis we made use of the following observations.
CRIRES data of \object{GJ~1002} and GJ~1224 were obtained during several nights in July 2007, 
as well as in May and August 2009 under programme IDs 079.D-0357 and 083.D-0124. 
The nominal resolving power was $R\approx100\,000$.
Data reduction made use of the ESOREX pipeline for CRIRES and a
custom-made IDL pipeline.
We find the $S/N$ at the continuum level of most parts of the final 
spectrum of GJ~1002 to exceed $200$. The $S/N$ of GJ~1224 is lower
and less homogeneous: we find it at the continuum level
to be $\approx170$ for the FeH regions used in this study
but only $\approx70$ for the metal lines longer than $1\mu$m.

The data for AD~Leo, YZ~CMi, and EV~Lac were taken with HIRES at Keck~I during three observing runs
on 2005 March~1, 2005 August~14 and~15, and 2005 December~18. Expected resolution is $R\approx31\,000$ for
AD~Leo and EV~Lac with $S/N=100$ and $S/N=140$ respectively; and $R\approx70\,000$, $S/N=70$
for YZ~CMi \citep[see][for further details]{r-and-b-2007}.

Additionally, as template spectra we used the FEROS (La Silla, Chille) data of two non-active M-dwarfs
\object{Gl~682}~(M4.0) and \object{LHS~337}~(M4.5). The data were taken on 2006 July~25 for the former and on 2007 March~4
for the former respectively. The resolution is $R=48\,000$.

\section{Methods}
\label{sec:methods}

\subsection{Input line lists and synthetic spectra}
In the attempt to model the Zeeman splitting in atmospheres of M-dwarfs we made use of both atomic
and molecular spectra available to us. Ideally, it is expected that this set of
lines contains both magnetically sensitive (for measuring field intensity) and insensitive (to fix atmospheric
parameters) lines. In the past several years, FeH lines of the Wing-Ford band ($F^4\,\Delta-X^4\,\Delta$ transitions,
$0.9-1\,\mu$m) were found to be an excellent diagnostics of the magnetic field in spectra of sunspots
and cooler M-dwarfs \citep[see, for example,][]{valenti2001,r-and-b-2006,r-and-b-2007,afram2007,afram2008,harrison-brown2008}.
The line list of FeH transitions and molecular constants were taken from \citet{dulick2003}\footnote{http://bernath.uwaterloo.ca/FeH/}.
We notice that the positions 
and strengths of some FeH lines are not accurate as provided 
by latter computations. Using spectra of non-magnetic M- 
dwarf GJ~1002, we tried to correct these lines in a way that they match 
observations and are consistent with other FeH lines which do 
not require any adjustments.  For this we identified as many as possible 
FeH lines in the observed spectra by hand and corrected their positions if necessary. 
The identification of the lines were confirmed by statistical means, cross-correlation 
techniques, and line intensities. 
For some FeH lines it was necessary to adjust their Einstein~A values to match the observations and thus
obtain a consistent fit between all of them, several strong atomic lines of \ion{Ti}{i}, and with the same model atmosphere
(see also Sect.~\ref{sec:general-notes}).
This is described in more detail in \citet{wende2010}.

The VALD (Vienna Atomic Line Database) was used as a source of atomic transitions \citep{vald1,vald2}.
We decreased the \loggf\ values of \ion{Ti}{i} $10\,610$~\AA\ and $10\,735$~\AA\ lines 
by $0.2$~dex to match observations of a non-active star GJ~1002. This ensured a consistent fit simultaneously 
with other strong Ti lines using same atmospheric parameters (see Sect.~\ref{sec:results}).

To compute synthetic spectra of the atomic and molecular lines in the magnetic
field we employed the \synmast\ code \citep{synthmag2007}. The code
represents an improved version of the \synthmag\ code described by
\citet{synthmag}. It solves the polarized radiative transfer equation for a
given model atmosphere, atomic and molecular line lists, and magnetic field
parameters. The code treats simultaneously thousands of blended absorption
lines, taking into account their individual magnetic splitting patterns, which
can be computed for the Zeeman or the Paschen-Back regime. \synmast\ provides 
local four Stokes-parameter spectra for a number of angles between the
surface normal and the line of sight (seven by default). These local spectra are
convolved with appropriate rotational, macroturbulent, and instrumental profiles
and then combined to simulate the stellar flux profiles.

A simple magnetic field model, described with a small number of free
parameters, was adopted in our calculations. The field is homogeneous in the
stellar reference frame and is specified by the three vector components: radial 
$\br$, meridional $\bm$, and azimuthal $\ba$, reckoned in the spherical coordinate
system whose polar axis coincides with the line of sight. Then, the two field
components relevant for calculating the Stokes $I$ profiles are given by
\begin{equation}
B_{\rm l} = \br\cos{\theta} - \ba\sin{\theta}
\end{equation}
for the line of sight component and
\begin{equation}
B_{\rm t} = \left[ (\br\sin{\theta} + \ba\cos{\theta})^2 + \bm^2 \right]^{1/2}
\end{equation}
for the transverse component. In practice it is sufficient to adjust $\br$ and
$\bm$, keeping $\ba$ zero.

This approximation of the magnetic field structure is undoubtedly simplistic
and unsuitable for describing phase-dependent four Stokes-parameter profiles
of magnetic M~dwarfs. Nevertheless, as proved by the previous studies of 
Zeeman-sensitive lines in cool stars \citep{jk-valenti1996,jk2007},
it is sufficient for modelling unpolarized spectra of M~dwarfs and T~Tauri stars
with strong fields.


Model atmospheres are from the recent MARCS grid\footnote{http://marcs.astro.uu.se} \citep{marcs}.

\subsection{Molecular Zeeman effect}
In order to analyse the magnetic field through the spectra synthesis it is necessary to know
the Land\'e g-factors of the upper and lower levels of a particular molecular transition.
These g-factors are defined by the corresponding wave-functions of individual states, and can be
in principle computed by the construction of an effective Hamiltonian for a given system of levels.
In diatomic molecules, there are several limiting cases of the level splitting 
defined by the strength of coupling between electronic orbital angular momentum $\mathbf{L}$ and electronic spin $\mathbf{S}$,
$\mathbf{L}$ and $\mathbf{S}$ to the internuclear axis, and $\mathbf{L}$ and $\mathbf{S}$ to the total
angular momentum $\mathbf{J}$ \citep{herzberg1950}.
In the present investigation two limiting cases are of particular interest.
Strong coupling of $\mathbf{S}$ and $\mathbf{L}$ to the internuclear axis (and thus their weak coupling with nuclear rotation)
is called \hunda. A weak coupling of $\mathbf{S}$ with internuclear axis is described by \hundb. Consequently,
states with splitting patterns between these extreme cases are called intermediate between pure Hund's cases (a) and (b).
Important to us, simple analytical expressions for g-factors
can only be obtained in pure (a) and (b) cases. Sad but true, as stated in \citet{berdyugina2002}, 
the lines of FeH of Wing-Ford band exhibit splitting,
which is in most cases intermediate between pure Hund's (a) and (b) 
and which is not trivial to treat both theoretically and numerically.
This implies that to compute g-factors,
one has to construct a kind of effective Hamiltonian, which would provide a realistic estimate
for the level energies affected by the external magnetic field. Usually this is done by
representing the total (effective) Hamiltonian as a sum of the unperturbed part, which describes
energies of Zeeman levels as they undergo the transition between Hund's cases, 
and the part which describes an interaction with the external
magnetic field. A detailed description of this procedure can be found in \citet{berdyugina2002} and \citet{ramos2006}.

In the present work, we implement numerical libraries from the MZL (Molecular Zeeman Library)
package originally written by B.~Leroy \citep{mzl}, and adopted
for the particular case of FeH. The MZL is a collection of routines for computing the Zeeman effect in diatomic molecules,
and it contains the complete physics of pure and intermediate Hund's cases presented in \citet{berdyugina2002}. 
For the atomic lines the g-factors were directly extracted from VALD.

\section{Results}
\label{sec:results}

\subsection{Atomic lines in sunspot spectra}

\begin{figure}
\includegraphics[width=\hsize]{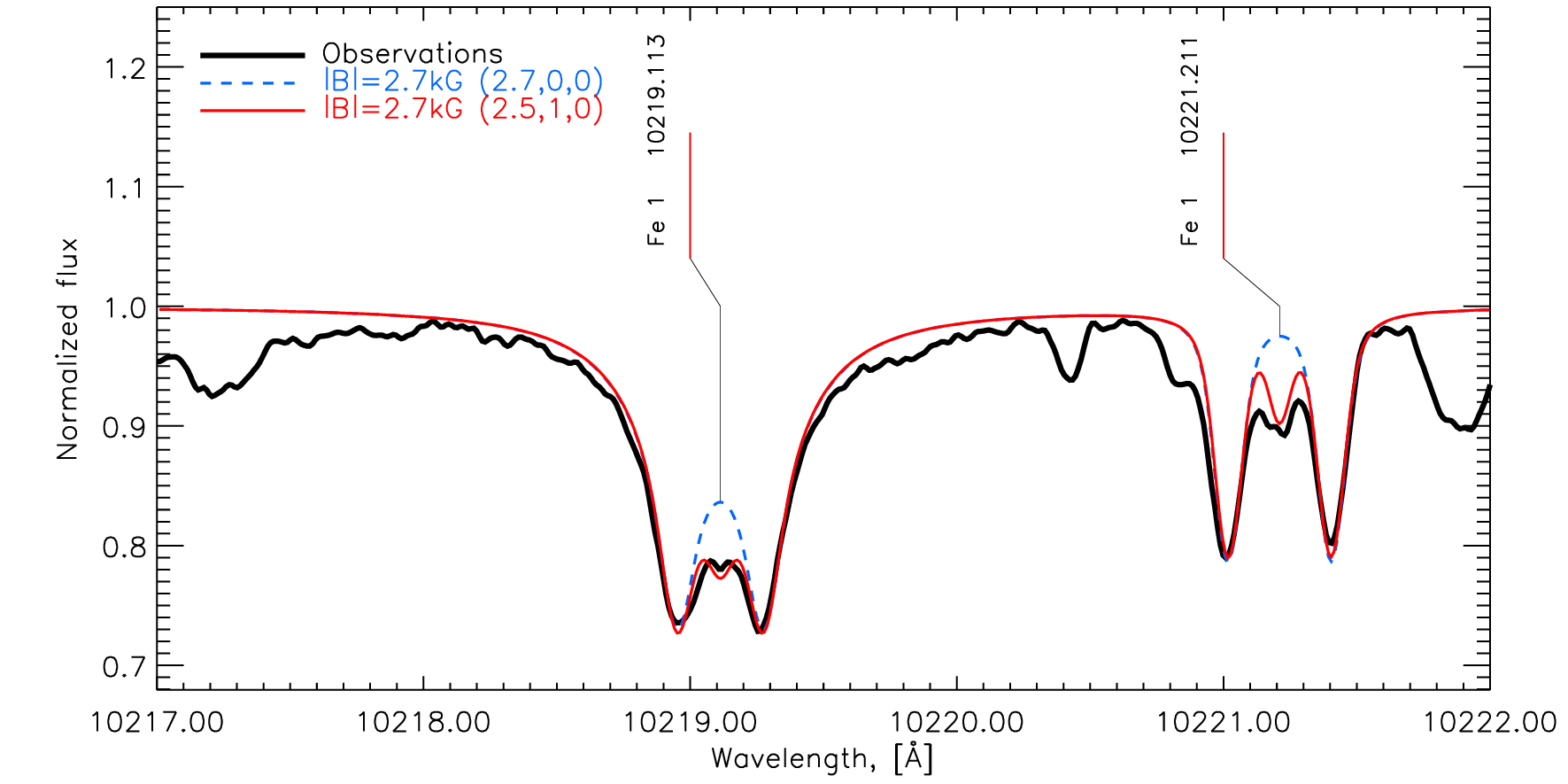}
\includegraphics[width=\hsize]{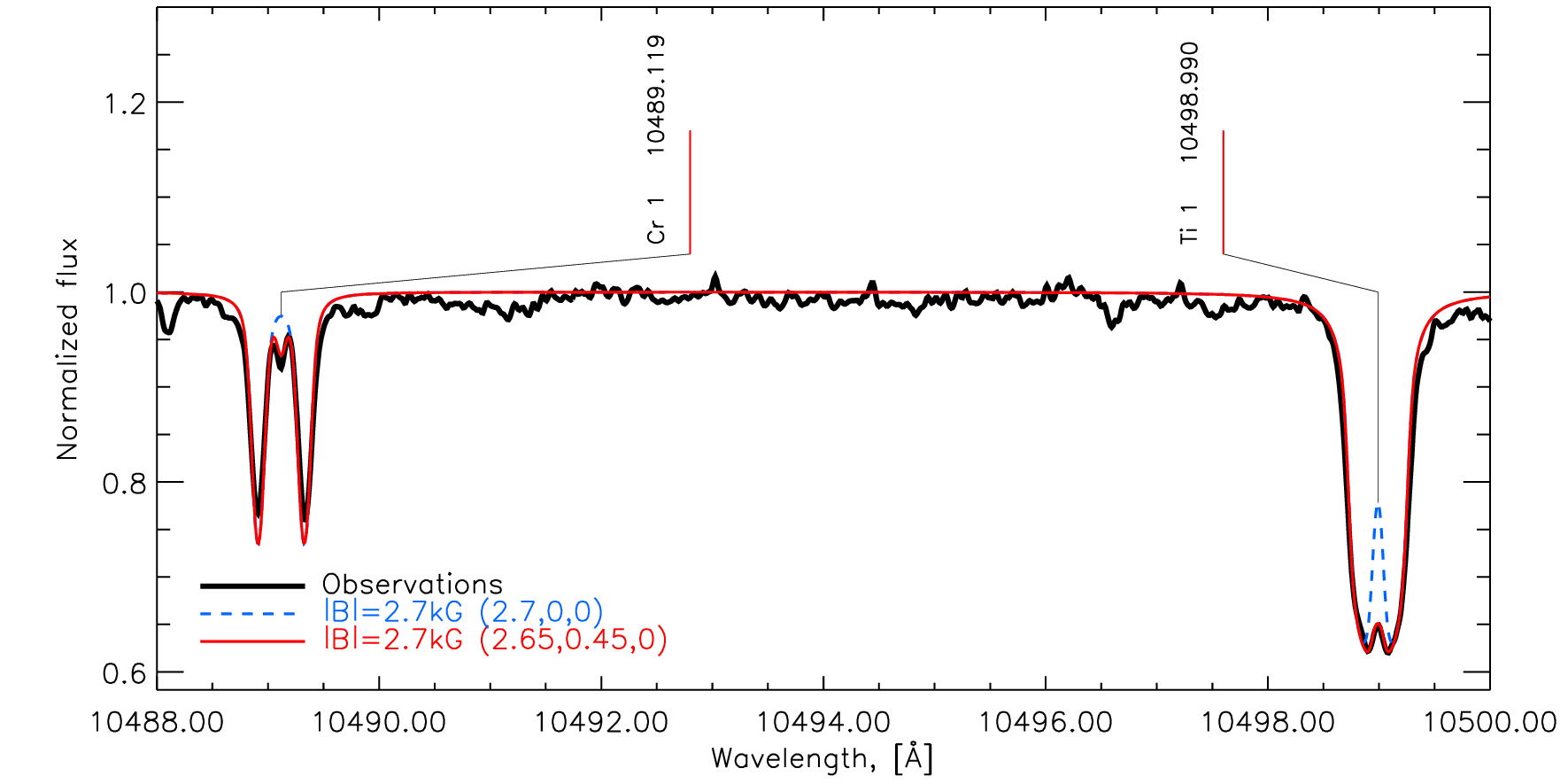}
\caption{Observed and predicted spectra of a Sun spot. Theoretical computations are shown for two
magnetic field geometries: purely radial ($\br,0,0$) and with horizontal contribution ($\br,\bm,0$)
(see legends on individual plots). Wavelengths are in vacuum.}
\label{fig:spot-atoms}
\end{figure}

Before presenting results of the M-star spectra analysis, it is necessary to verify our methods through
observations of some standard star where both atomic and molecular lines can be seen simultaneously.
A sunspot spectrum is probably the only trustworthy data source in this regard because 
a) the temperature inside a spot is still hot enough to see strong unblended atomic and FeH lines and 
b) very high-resolution and S/N observations are available. We thus made use of an umbral spectrum
from \citet{wallace1998}. They also derived the magnetic field intensity $\b=2.7$~kG.
Our fit to the atomic lines (mostly Fe, Ti, and Cr) in the range $9800-10\,800$\,\AA\ also 
suggests a field intensity $\b=2.7$~kG (model atmosphere with
$\teff=4000$~K and solar abundances from \citet{asplund2005}). Note that to simulate an intensity spectrum inside the spot,
the radiative transfer was solved for the pencil of radiation having $\mu=1$ ($\mu=\cos\theta$, $\theta$~--~angle
between normal to the surface and the line-of-sight). The observed line profiles can be fitted assuming 
the geometry of the magnetic field with dominated radial component, but an additional non-zero horizontal component is required,
whose intensity can actually vary from line to line.
As an example, Fig.~\ref{fig:spot-atoms} illustrates theoretical fits for some selected atomic lines
under different assumptions of the field geometry. For a purely radial field, only $\sigma$-components
must be visible, but many lines still illustrate strong $\pi$-components. For instance, the fit to the \ion{Fe}{i}
lines in the upper plot of Fig.~\ref{fig:spot-atoms} require twice weaker meridional $\bm$ (or horizontal) field component,
while the \ion{Cr}{i} and \ion{Ti}{i} lines shown in the bottom plot need an even weaker $\bm$. 
Note that $\b$ is always $2.7$~kG, as defined from the analysis of the position of Zeeman components,
while the same filed modulus can be constructed by assuming different magnetic field geometries, i.e. by adjusting 
the intensities of vector components.

\subsection{Land\'e factors of FeH lines}

\begin{figure*}[ht]
\includegraphics[width=\hsize]{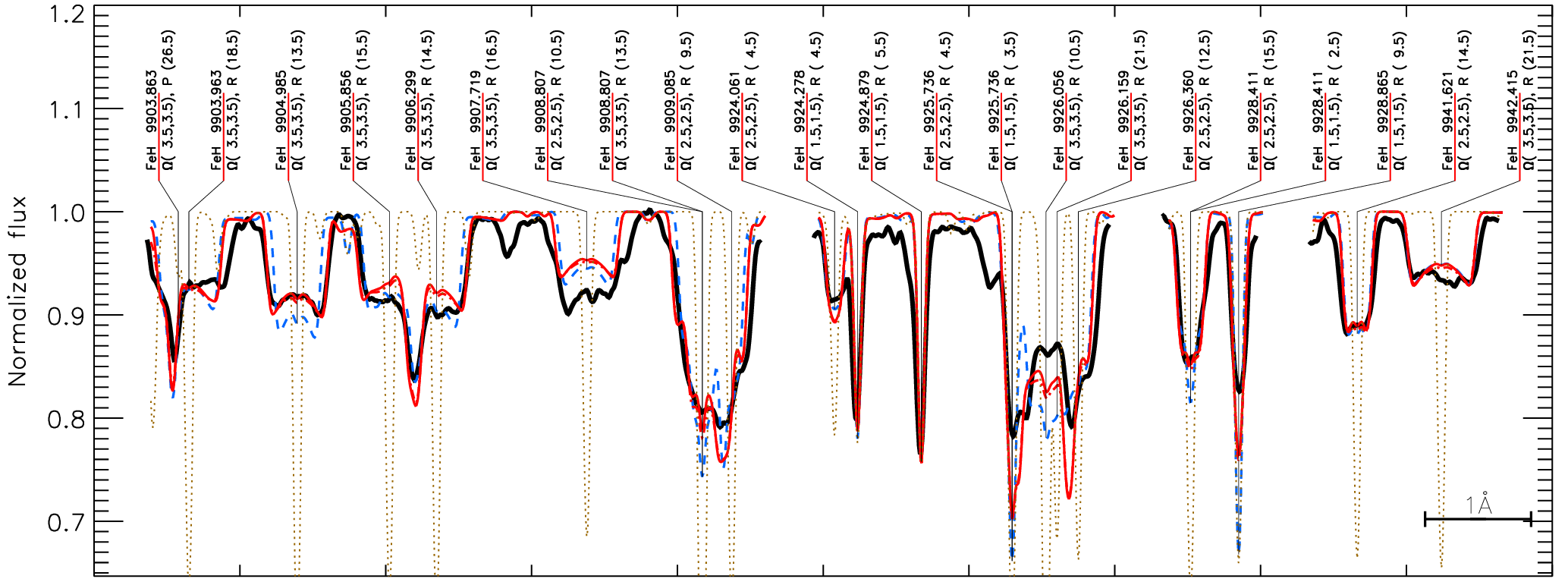}
\includegraphics[width=\hsize]{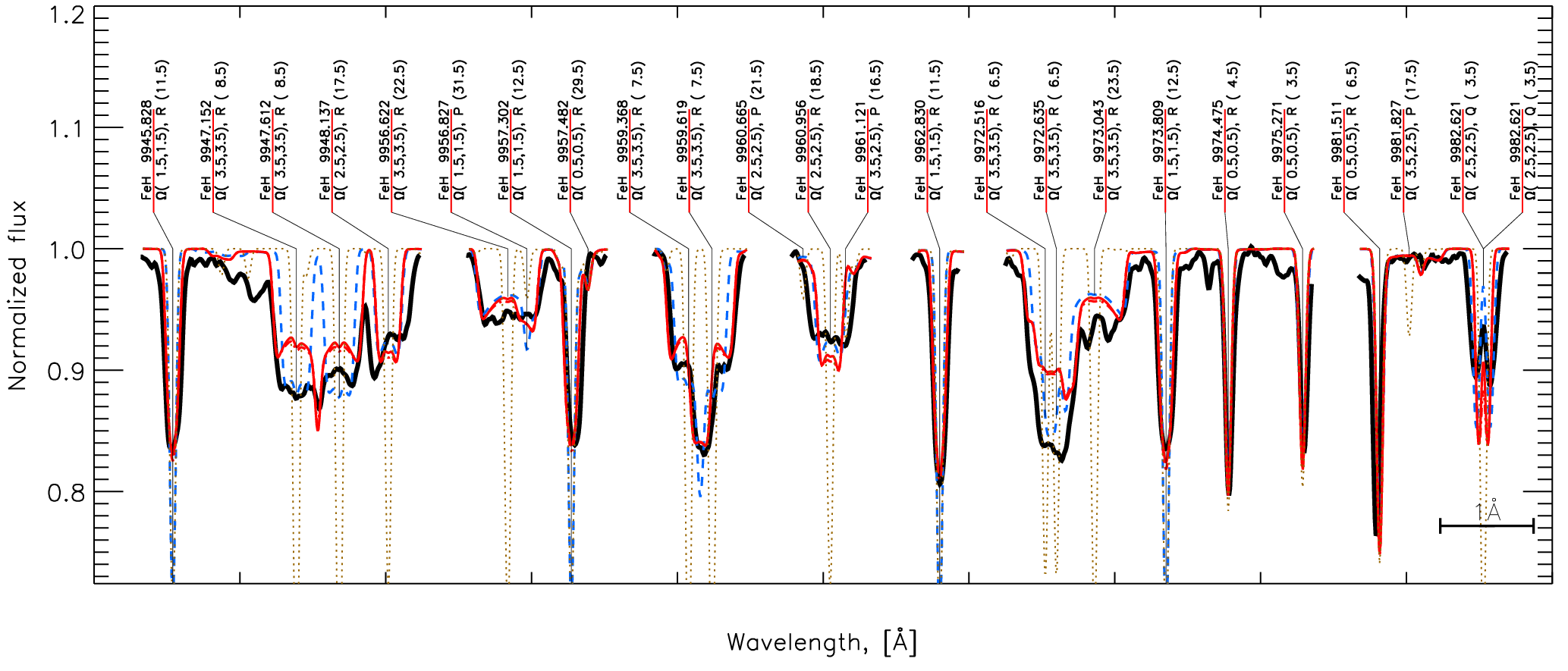}
\caption{Comparison between observed and theoretical sunspot spectra in selected regions of FeH transitions. 
Blue dashed line~--~best-fitted g-factor from \citet{afram2008}, red solid line~--~calculations with MZL library; 
both with a purely radial field of $\bv=(2.7,0,0)$~kG. Red dash-dotted line~--~ synthetic spectra accounted for the horizontal field 
component $\bv=(2.5,1,0)$~kG (hardly seen in the figure, coincides with red solid line),
brown dotted line~--~zero-field spectrum.
Labels over lines indicate their central wavelengths, omegas of lower and upper states (in brackets), branch, and the $J$-number
of the lower state (in brackets). Wavelengths are in vacuum.}
\label{fig:spot-feh}
\end{figure*}

As noted above, Wing-Ford lines of FeH are mostly formed in intermediate Hund's case, which makes it difficult to 
accurately predict Zeeman patterns of respective states.
Analytically, the type of a splitting can be estimated
via the analysis of spin-orbit and rotational constants $Y=|A_{\rm \upsilon}|/B_{\rm \upsilon}$: if $Y\gg J(J+1)$ when an approximation
of \hunda\ is valid, and \hundb\ otherwise \citep{herzberg1950}. 
Using an empirical analysis of laboratory data, \citet{harrison-brown2008} determined
g-factors for the upper and lower levels for a number of states with $\Omega=7/2, 5/2, 3/2$. Their Fig.~2
clearly shows a deviation from pure Hund's cases for upper levels with $J>4.5$. Almost at the same time, \citet{afram2008}
presented polynomial best-fit g-factors based on an accurate and extensive analysis of sunspot spectra.
These approaches clearly illustrated the main problem: it is impossible to fit all FeH lines simultaneously 
with the modern theory of the intermediate Hund's case. Instead, a good fit can be obtained only by combining 
theoretical and empirical approaches.

Using the MZL library and sunspot spectra we tried to compute theoretical g-factors for all lines under the condition
that the resulting Zeeman patterns provide correct broadening of individual FeH lines in sunspot spectra. 
In general, we find that the intermediate case (with its present treatment in MZL) is a good approximation if 
($l$~--~lower, $u$~--~upper states)
\begin{enumerate}
\item
$\Omega_l=0.5$
\item
$\Omega_{l \; \mathrm{or} \; u}\leq2.5$ and $3Y>J(J+1)$ for the P and Q branches
\item
$\Omega_{l \; \mathrm{and} \; u}=2.5$ and $5Y>J(J+1)$ for the R branch.
\end{enumerate}
For the rest of transitions the assumption of \hunda\ for the upper level 
and \hundb\ for lower level provide reasonable results, 
especially for transitions with $\Omega_{l  \; \mathrm{and} \; u}=3.5$.

\begin{figure*}[!ht]
\includegraphics[width=\hsize]{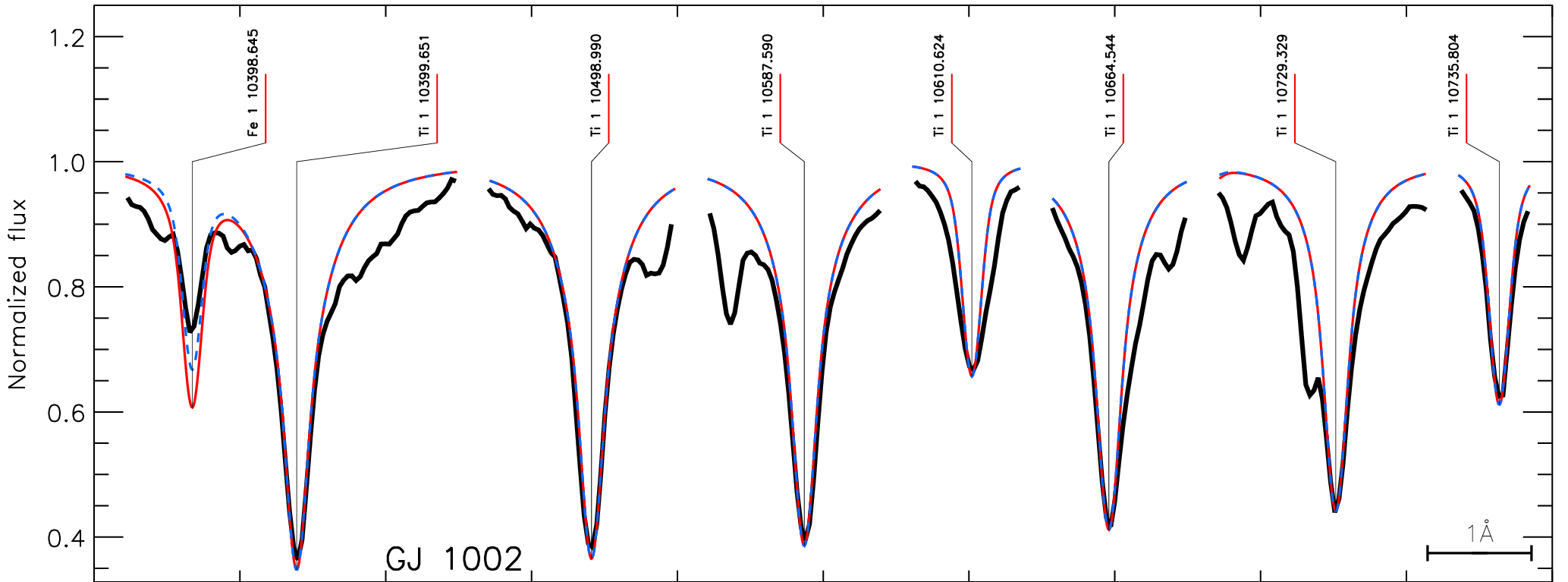}
\includegraphics[width=\hsize]{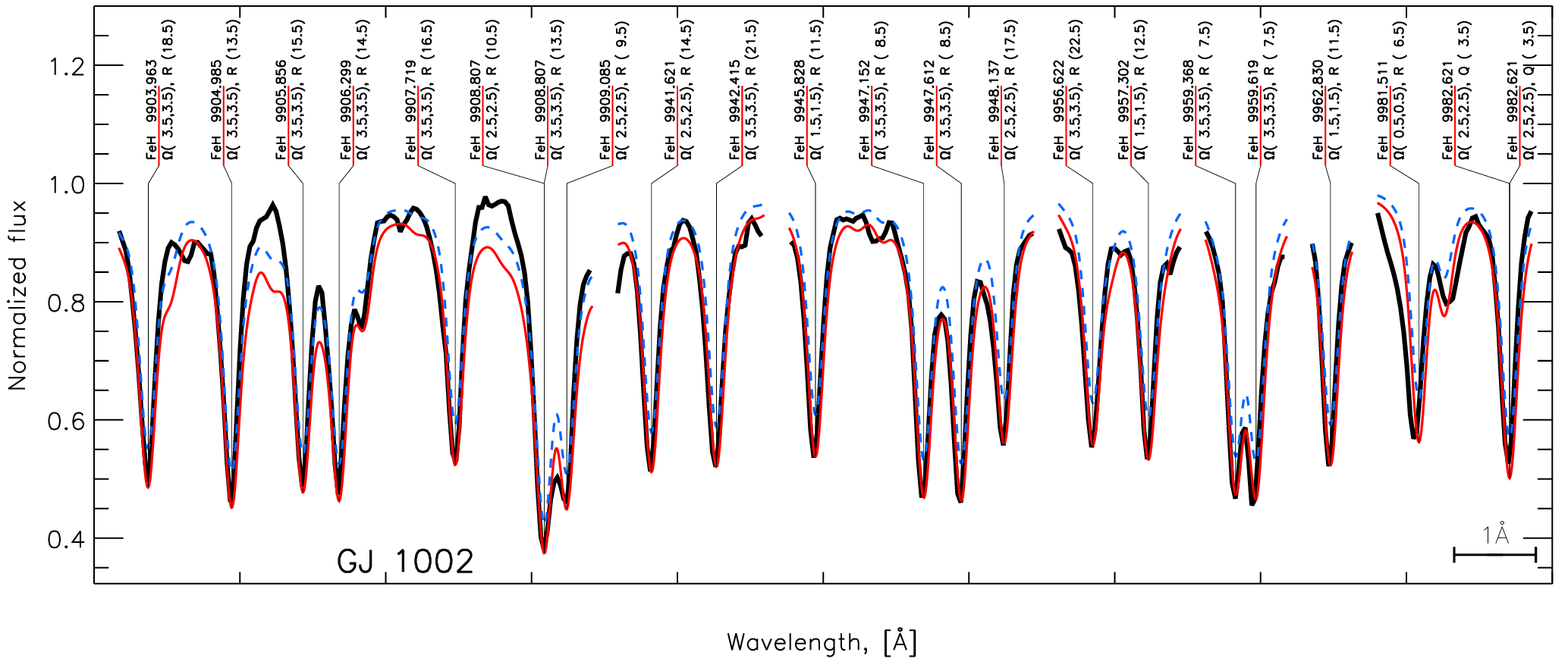}
\caption{Observed and predicted spectra of M$5.5$ dwarf GJ~1002 in \ion{Ti}{i} (\textit{upper plot})
and selected FeH lines (\textit{lower plot}).
Model parameters: $\teff=3100$~K, $\logg=5.0$, [M/H]$=0.0$, \vsini$=2.5$~\kms. Observations are shown by thick solid line. 
Theoretical spectra were computed assuming $\varepsilon_{\rm Fe}=-4.37$ (red solid line) and $\varepsilon_{\rm Fe}=-4.59$ (blue dashed line) respectively.
Wavelengths are in vacuum.}
\label{fig:gj1002}
\end{figure*}

Figure~\ref{fig:spot-feh} illustrates a comparison between best-fitted g-factor from \citet{afram2008} and our calculation for
some selected lines in the $9900-10\,000$~\AA\ region. Note that apart from the figures shown in \citet{afram2008}, we did not make an attempt
of correcting the theoretical spectra, i.e. no filling factors were applied. The model atmosphere and field intensity are the same as
determined previously from the metallic line spectra. 
From Fig.~\ref{fig:spot-feh} it is obvious
that there is in general a good agreement between our calculation and g-factors from \citet{afram2008}. Still, the discrepancy between the two calculations
is found for low omega R-branch lines like FeH $9945$~\AA, $9962$~\AA, etc., which indeed show splitting closer to pure Hund's cases.
Applying the intermediate case allows us to predict the splitting of FeH $9982$~\AA\ and the magnetic insensitive $9981$~\AA\ line.
On the other hand, there is systematic difference seen for some lines ($9904.98$~\AA, $9947$~\AA, etc.), 
where our calculations predict broader and shallow line profiles then those given in \citet{afram2008}. 
Even though we succeeded well enough to fit the width of the observed lines with the same field of $\b=2.7$~kG derived previously from atomic lines,
this cannot be judged to be more physical though until new improvements in the theoretical description of the intermediate case
will become available.
In spite of atomic lines, we find little sensitivity of the FeH lines to the magnetic field geometry
changing from purely radial to the additional contribution of the horizontal component. 
What finally matters is the total field intensity. 
Note, however, that it is still possible to distinguish between the \textit{dominated} horizontal or radial field components 
via the detailed analysis of magnetically broadened FeH lines.

Thus, applying different Hund's cases based on transition quantum numbers allows us to predict the Zeeman broadening
of certain FeH lines accurately and consistently (i.e. with the same field strength) with the atomic lines.

\subsection{Magnetic field of selected M-dwarfs}
\subsubsection{General notes}
\label{sec:general-notes}

A comparison of theoretical and observed sunspot spectra of FeH allows us to select lines with more or less accurately predicted Zeeman patterns
and to measure magnetic fields in cooler stars. Below we present results for a few selected M-dwarfs, for which previous attempts
revealed a presence of strong (up to $4$~kG) effective magnetic fields \citep[see][]{r-and-b-2007}. 


In cool atmospheres, the van der Waals constant is one of the most important broadening mechanisms.
For FeH, there are no theoretical or laboratory measurements
of broadening constants available. We thus checked the spectra of FeH in the atmosphere of the non-magnetic M$5.5$ star 
GJ~1002 and found that the classical van der Waals constant \citep[see][]{gray} used in \synmast\ must be increased 
by a factor of $3.5$ to satisfactorily match the profiles of FeH lines. A model atmosphere with 
$\teff=3100$~K, $\logg=5$, [M/H]=$0.0$, and \vsini$=2.5$~\kms\ 
can fit strong \ion{Ti}{i} lines in the region $10\,300-10\,800$~\AA, as shown in Fig.~\ref{fig:gj1002}.
However, using the solar abundance of iron $\varepsilon_{\rm Fe}=-4.59$ ($\varepsilon_{\rm Fe}=\log(N_{\rm Fe}/N_{\rm total})$)
results in FeH lines that are systematically too weak. Decreasing \vsini\ to $1$~\kms\ may solve this problem, 
but the cores of titanium lines are then impossible to fit.
Because the damping constant and \loggf's values for the considered Ti lines are known from 
more accurate calculations (which are based on observed energy levels  as presented in \citet{gfiron-ti}), 
we used them primarily for FeH lines as indicators 
of accurate model parameters and \vsini. The adjusted iron abundance is close to its previous solar value $\varepsilon_{\rm Fe}=-4.37$ \citep{grevesse1989}. 
Note that this, of course, in no way sets the correct metallicity of the star. But the lack of
weakly blended Fe lines and well-known problems with continuum normalization 
of small spectral regions forbids more definite results. 
One should not forget that measuring magnetic fields
mostly relies on the relative line broadening caused by Zeeman splitting, and thus the issue of metallicity is not that
important in the present investigation.

There is a principle difference between modelling spectra of a sunspot and distant stars. 
In the former case the observer receives an intensity spectrum. Because of high spatial resolution, it is natural to assume 
that all individual rays of light propagate along the line-of-sight. This allows one to solve the radiative transfer problem
only for one angle $\mu=1$. For distant stars, however, the surface integrated intensity
is what is observed by our instruments. Because the geometry of the magnetic field is now a function of surface coordinates,
we define the magnetic field components  (radial $\br$, meridional $\bm$, azimuthal $\ba$) at the centre of the stellar disc. 
The intensities of these components are then modified according to the local $\mu$ and $\phi$ spherical angles
on the stellar surface. The flux visible to the observer is obtained by angle integration of
local specific intensities. For instance, assuming a non-zero $\br$ and zero $\bm$ and $\ba$, i.e. $\bv=(\br,0,0)$,
may correspond to the dipolar-dominated field geometry with the magnetic pole at the centre of the disc and the magnetic axis
pointing along the line-of-sight.

Below we examine our approach of measuring magnetic field in atmospheres of selected M-dwarfs.

\begin{figure*}
\includegraphics[width=0.5\hsize]{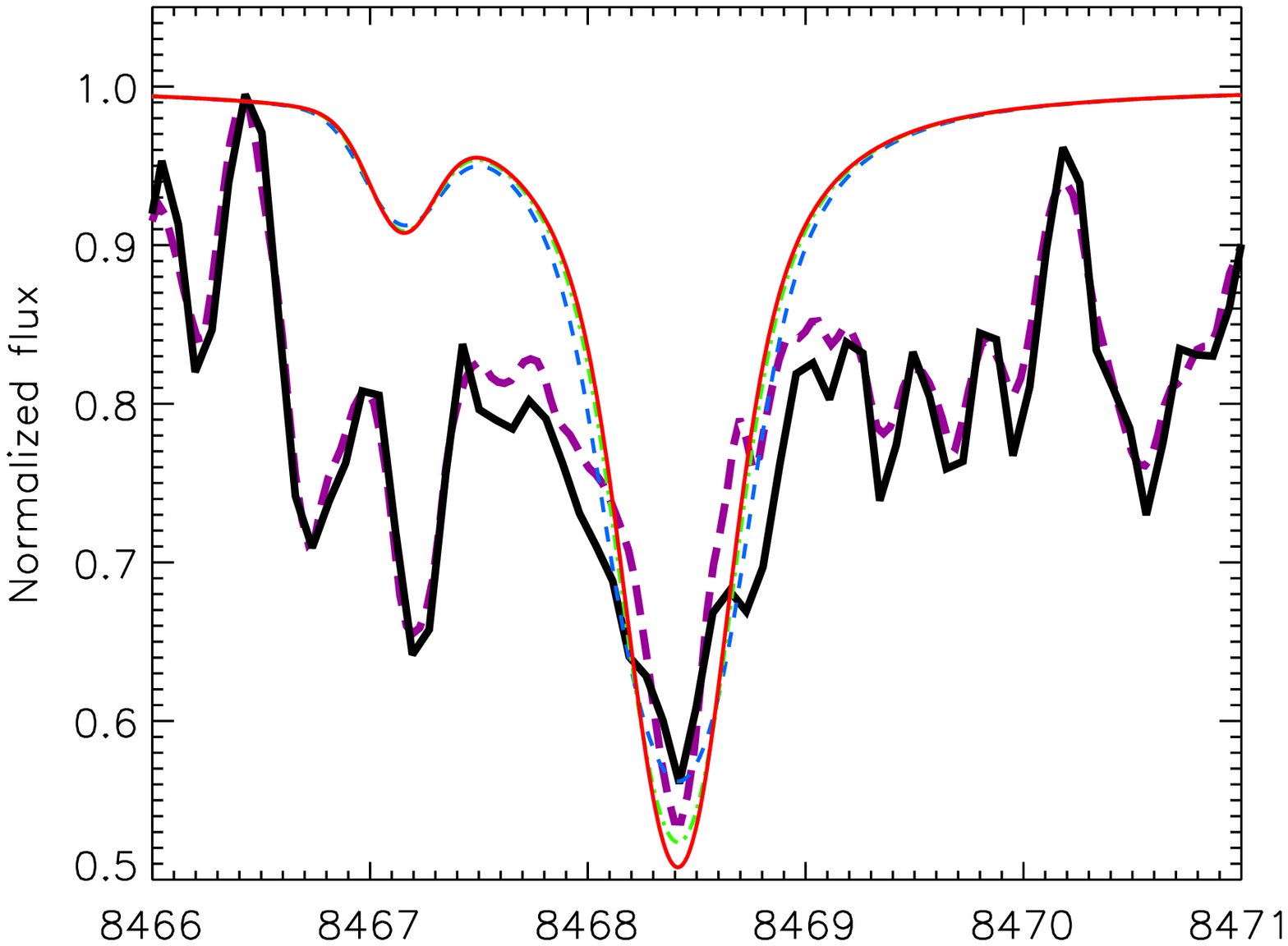}
\includegraphics[width=0.5\hsize]{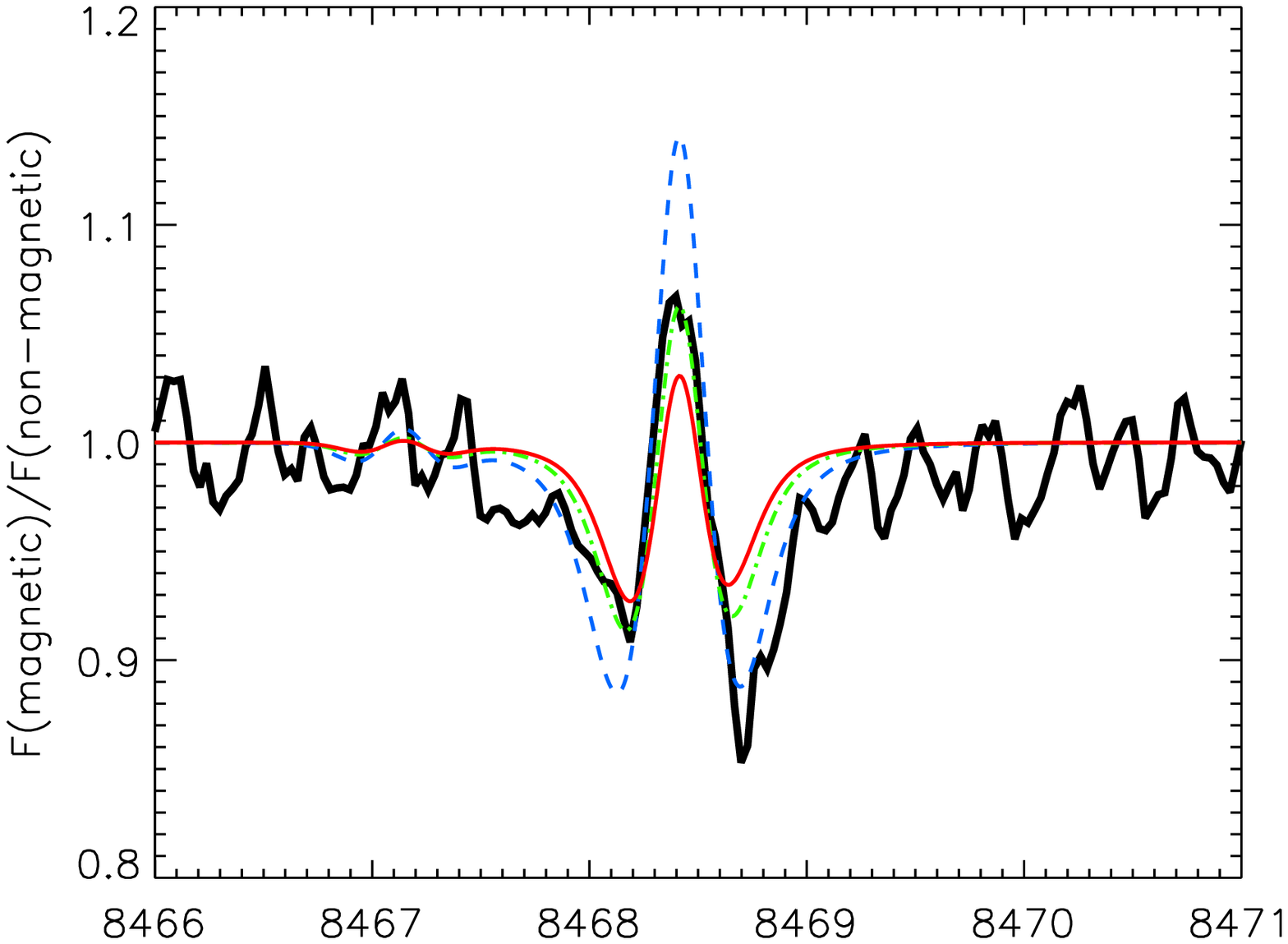}\\\\
\includegraphics[width=\hsize]{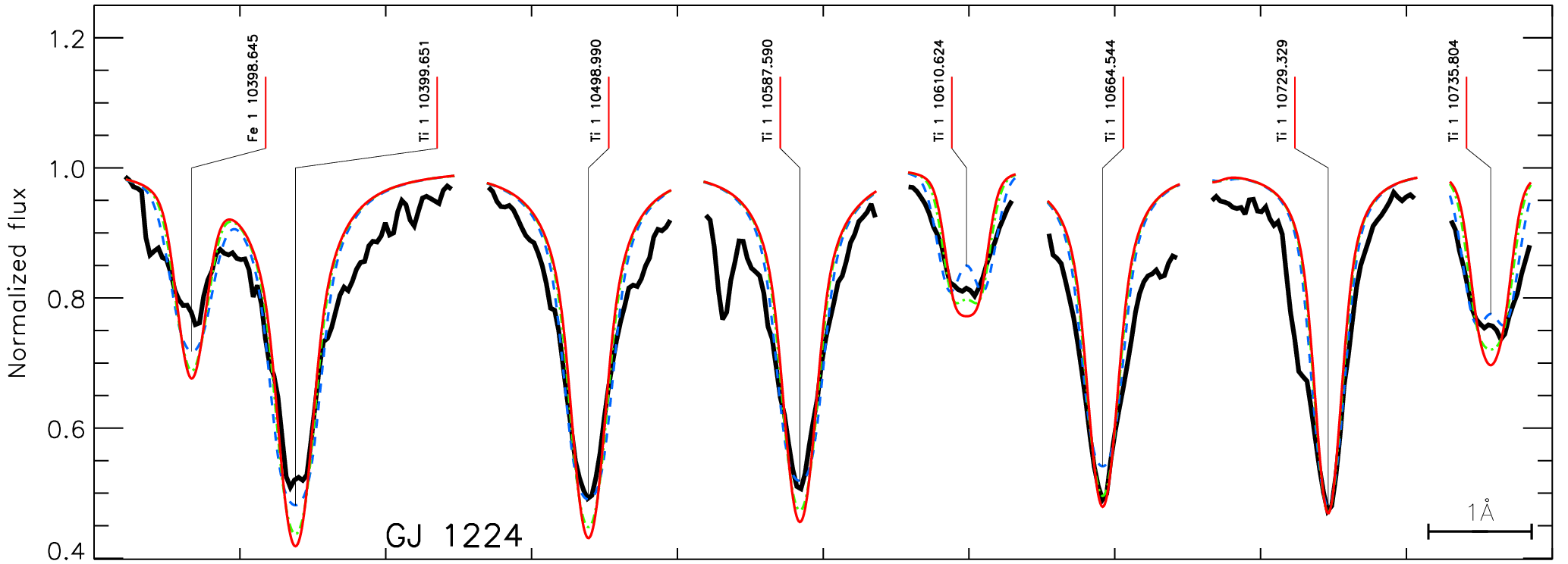}
\includegraphics[width=\hsize]{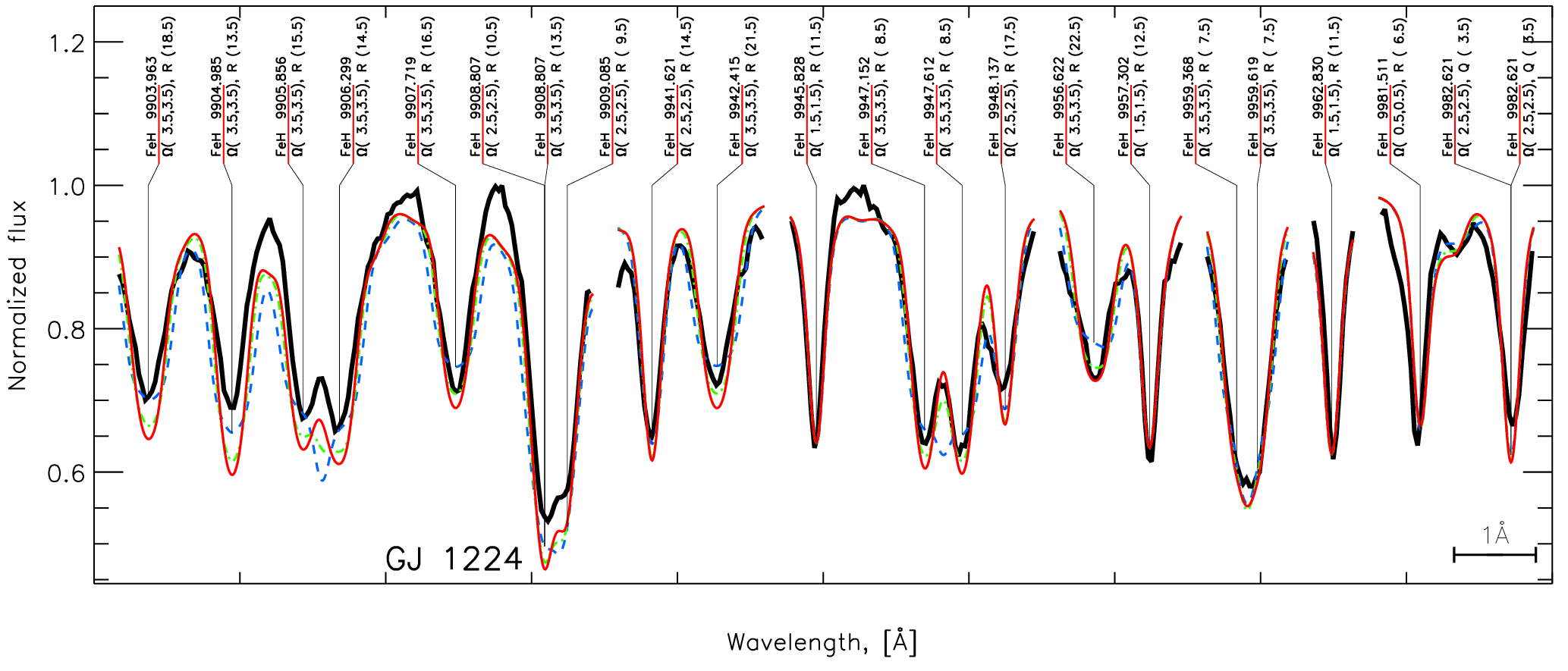}
\caption{Comparison between observed and theoretical spectra of the M$4.5$ dwarf GJ~1224. 
\textbf{Upper panel:} observed and calculated spectra at the \ion{Fe}{i}~$8468$~\AA\ line 
(\textit{left plot}) and the ratio between corresponding magnetic and non-magnetic spectra (\textit{right plot}). 
Thick solid line~--~observations of GJ~1224, violet long-dashed line~--~observations of inactive LHS~337.
\textbf{Middle and lower panels:} same as in Fig.~\ref{fig:gj1002}.
Model parameters: $\teff=3200$~K, $\logg=5.0$, $\varepsilon_{\rm Fe}=-4.45$, \vsini$=3$~\kms. Thick solid line~--~observations,
blue dashed line~--~ $\bv=(2.7,0,0)$~kG, green dash-dotted~--~$\bv=(2,0,0)$~kG, red solid line~--~ $\bv=(1.7,0,0)$~kG.
Wavelengths are in air for upper plots and in vacuum for the middle and lower plots.}
\label{fig:gj1224}
\end{figure*}

\subsubsection{GJ~1224}

The source GJ~1224 is an active  M$4.5$ dwarf for which previous attempts to measure its magnetic field 
resulted in $\btimesf=2.7$~kG \citep{r-and-b-2007}.
An analysis of the magnetic insensitive FeH lines and \ion{Ti}{i} $10\,729.3$~\AA\ line suggested $\teff=3200$~K, $\logg=5.0$ and 
\vsini$=3$~\kms\ under assumption of solar Ti abundances. We found it necessary to slightly increase the Fe abundance 
to $\varepsilon_{\rm Fe}=-4.45$ to match the magnetically insensitive FeH lines.

The upper plot of Fig.~\ref{fig:gj1224} compares the observed and predicted spectra of the \ion{Fe}{i} $8468$~\AA\ line,
which was previously used by \citet{jk-valenti2000} to estimate the magnetic field in a number of M-dwarfs. 
Here we made use of low resolution ($R=31\,000$) Keck observations.
There is a general difficulty of measuring the magnetic field strength from this line though because of heavy blending by TiO. 
For instance, the absorption feature in the red wing of the line (at $8468.7$~\AA)
is actually a TiO line which thus affects the Zeeman red-shifted $\sigma$-components of the \ion{Fe}{i} $8468$~\AA, whose
position is then difficult to estimate with high precision. Therefore, following \citet{jk-valenti2000}, 
we tried to investigate a relative line intensity by dividing the spectra of GJ~1224 by spectra of a non-magnetic M dwarf LHS~337, 
which has the same or very similar spectral type. Resulting profiles are shown on the right upper panel of Fig.~\ref{fig:gj1224}.
The field of $2.7$~kG can fit the slope of the red wing of the residual, but a weaker field of about $2$~kG is required for the
blue wing of the line. This discrepancy is likely due to a relatively low quality of the data and strong blending by TiO.
In addition, this relative analysis, and especially the strength of the deep features at both sides of the line centre (see right
upper plot of Fig.~\ref{fig:gj1224}), implicitly assumes similar iron abundances for both active and inactive stars,
which may not be the case.

The middle and lower plots of the Fig.~\ref{fig:gj1224} illustrate the fit to the same set of Ti and FeH lines as performed for 
the non-magnetic GJ~1002 assuming different magnetic field intensities. 
There are several things to note. First of all, it is impossible to fit the cores of magnetic-sensitive Ti lines with
any field geometry and intensity. Fields larger than $2$~kG result in cores that are too wide to be observed.
Inversely, using $\b\leqslant2$~kG allows us to obtain a reasonable fit to the width of line cores, 
but the predicted central depths are too strong for some lines.
This behaviour is broken for the Ti~$10\,664.5$~\AA\ line, for which $\b=2$~kG and $\b=1.7$~kG provide a good fit, and
the Ti~$10\,610$~\AA\ line whose core can only be described by $\b=2$~kG. Another line at $10\,735$~\AA\ seems to point in the
direction of $\b\approx2.7$~kG, but the data quality at the red end of the spectra is poor and it is impossible
to draw accurate conclusions.

Extending this analysis to the FeH lines brings stronger constraints for the magnetic field intensity, as shown in the lower panel of
the Fig.~\ref{fig:gj1224}. In particular, magnetically sensitive lines such as FeH $9905$~\AA, $9906$~\AA, $9942$~\AA, and $9959$~\AA\ clearly 
point to the field modulus $\b\leqslant2$~kG. That the widths of these lines are well reproduced in the sunspot spectra 
(see Fig.~\ref{fig:spot-feh}) allows us to consider them as important indicators of the mean field intensity. In particular,
increasing $\b$ results in the appearance of the characteristic feature owing to the crossed $\sigma$-components of the two
FeH lines at $9906$~\AA. Overlaid, these components give rise to the absorption feature \textit{which is not seen in the observed spectra}.
Consequently, weaker fields are needed to keep these lines separated. Changing the magnetic field geometry by varying the strength
of the horizontal field components does not help to disable this feature: in our computations this is only possible 
with $1.7\leqslant\b\leqslant2$~kG, preferably with $\b\approx1.7$~kG. Finally, the same conclusions are reached in the analysis of the Keck
data available to us, but because of their lower resolution compared to that of CRIRES we did not include them in the plot.

\begin{figure*}[ht]
\includegraphics[width=0.5\hsize]{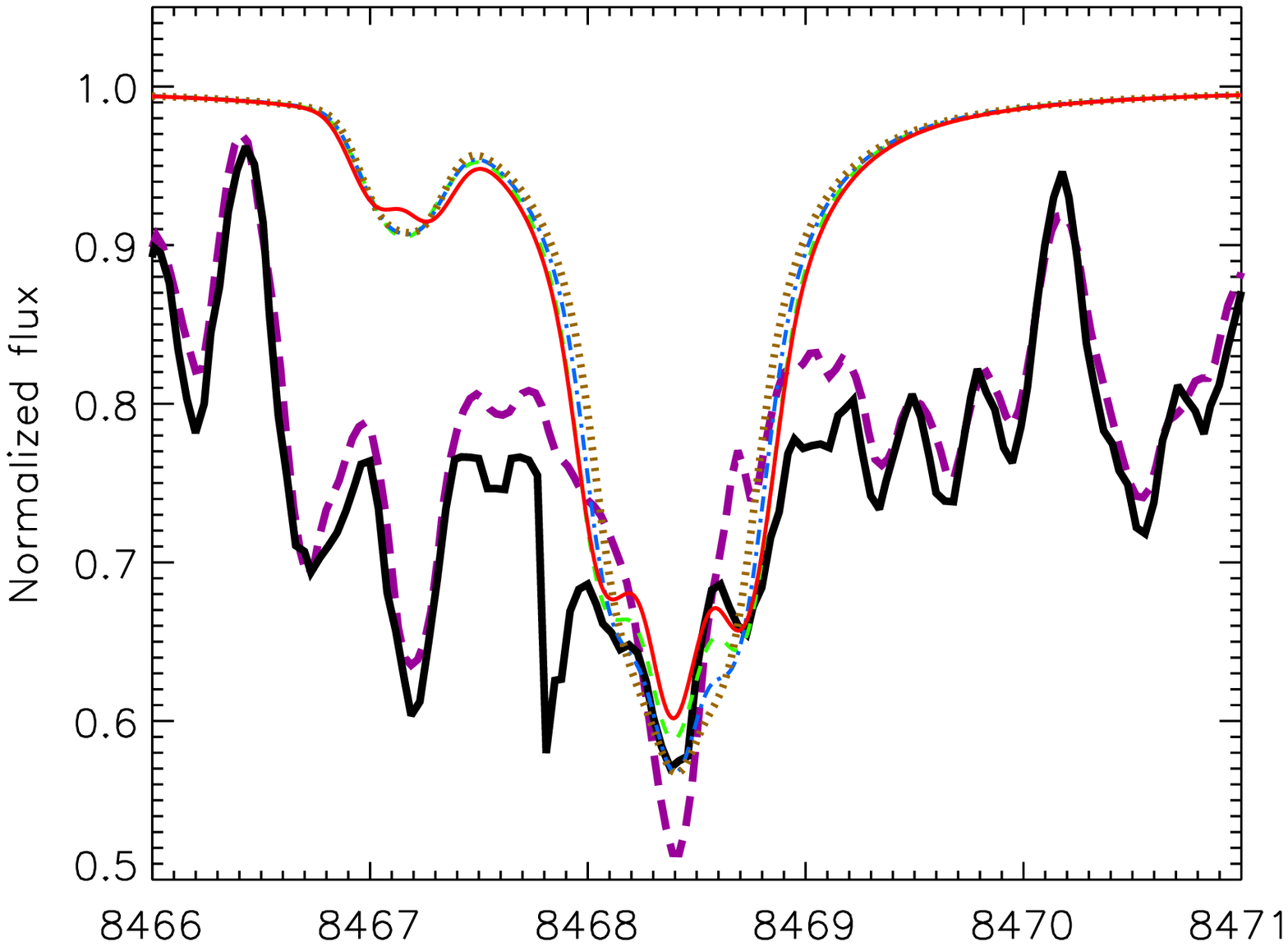}
\includegraphics[width=0.5\hsize]{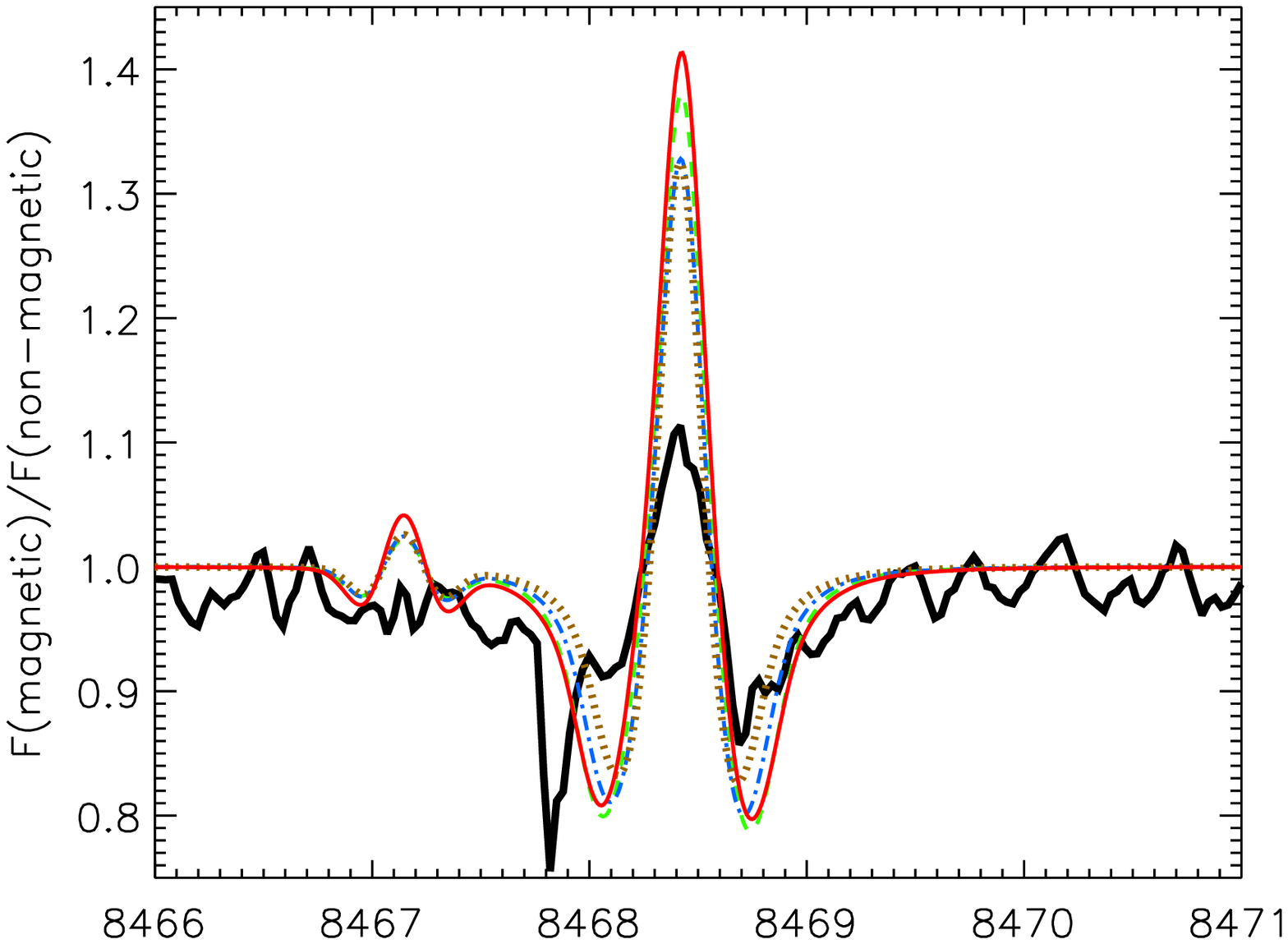}\\\\
\includegraphics[width=\hsize]{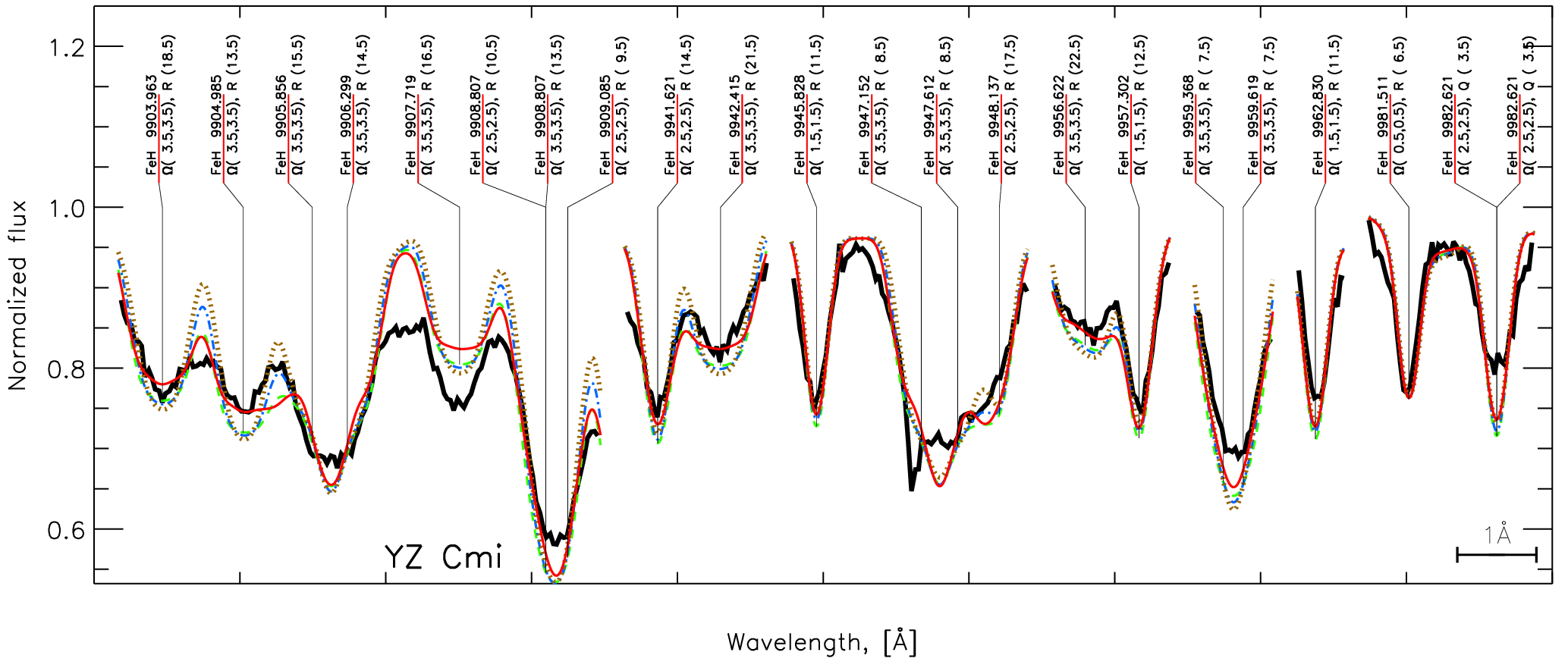}
\caption{Observed and predicted spectra of the M$4.5$ dwarf YZ~CMi.
Model parameters: $\teff=3300$~K, $\logg=5.0$, $\varepsilon_{\rm Fe}=-4.40$, \vsini$=5$~\kms. Thick solid line~--~observations,
violet long-dashed line (upper left plot only)~--~observations of inactive LHS~337,
green dashed~--~ $\bv=(0,4,0)$~kG, blue dash-dotted~--~$\bv=(2.5,2.5,0)$~kG, red solid~--~ $\bv=(4,0,0)$~kG,
dotted~--~$\bv=(3,0,0)$~kG. Wavelengths are in air for the upper plot and in vacuum for the lower.}
\label{fig:yzcmi}
\end{figure*}

\subsubsection{YZ CMi}

The source YZ~CMi (GJ~285) is a particular example because of the large $\btimesf > 3.9$~kG, which 
is the lower limit of an average field as derived by \citet{r-and-b-2007}.
Indeed, a $\b\approx4$~kG magnetic field seems to follow from the analysis of some atomic lines in the visual region of the
high-resolution ($R=70\,000$) Keck spectra. Figure~\ref{fig:yzcmi} illustrates theoretical calculations for
the \ion{Fe}{i} $8468$~\AA\ line in the same manner as was done for GJ~1224 (the adopted Fe abundance is $\varepsilon_{\rm Fe}=-4.40$).
For instance, a field of $\sim4$~kG is needed to place the $\sigma$-components
right where the TiO absorption feature is, but a somewhat weaker field of $3.5$~kG is also sufficient.
Unfortunately, a low signal-to-noise ratio of the observations brings only more uncertainty
in the analysis and makes it difficult to chose a unique field intensity.

The predicted profiles of the FeH lines are shown in the lower plot of Fig.~\ref{fig:yzcmi}. 
Assuming more or less accurate continuum normalization in the region $9904$~--~$9906$~\AA, the fields of $4$~kG
provide a better fit to the line depths, but the Zeeman broadening is then too strong.
The same holds true for FeH $9942$~\AA, $9956$~\AA.
A weaker field of $3.5$~kG with equal contributions from radial and horizontal
components $\bv=(2.5,2.5,0)$~kG is also a good approximation. Here again, as for the atomic lines, the low signal-to-noise
ratio prevents the detailed analysis of the line core regions that carry the most important information about the magnetic field modulus.
For instance, the width of the FeH lines in the $9904$~--~$9906$~\AA\ region are very well reproduced in the sunspot spectra. In this light
the field of YZ~CMi should be well below $4$~kG. Decreasing the Fe abundance could result in a stronger field, but this is not supported
by the magnetically insensitive FeH lines. Generally, the FeH lines point in the direction of a weaker field than previously reported. 
Less can be said about the geometry of the magnetic field, but a strong horizontal component can also be present. 
Higher S/N observations are strongly needed.

\begin{figure*}
\includegraphics[width=0.5\hsize]{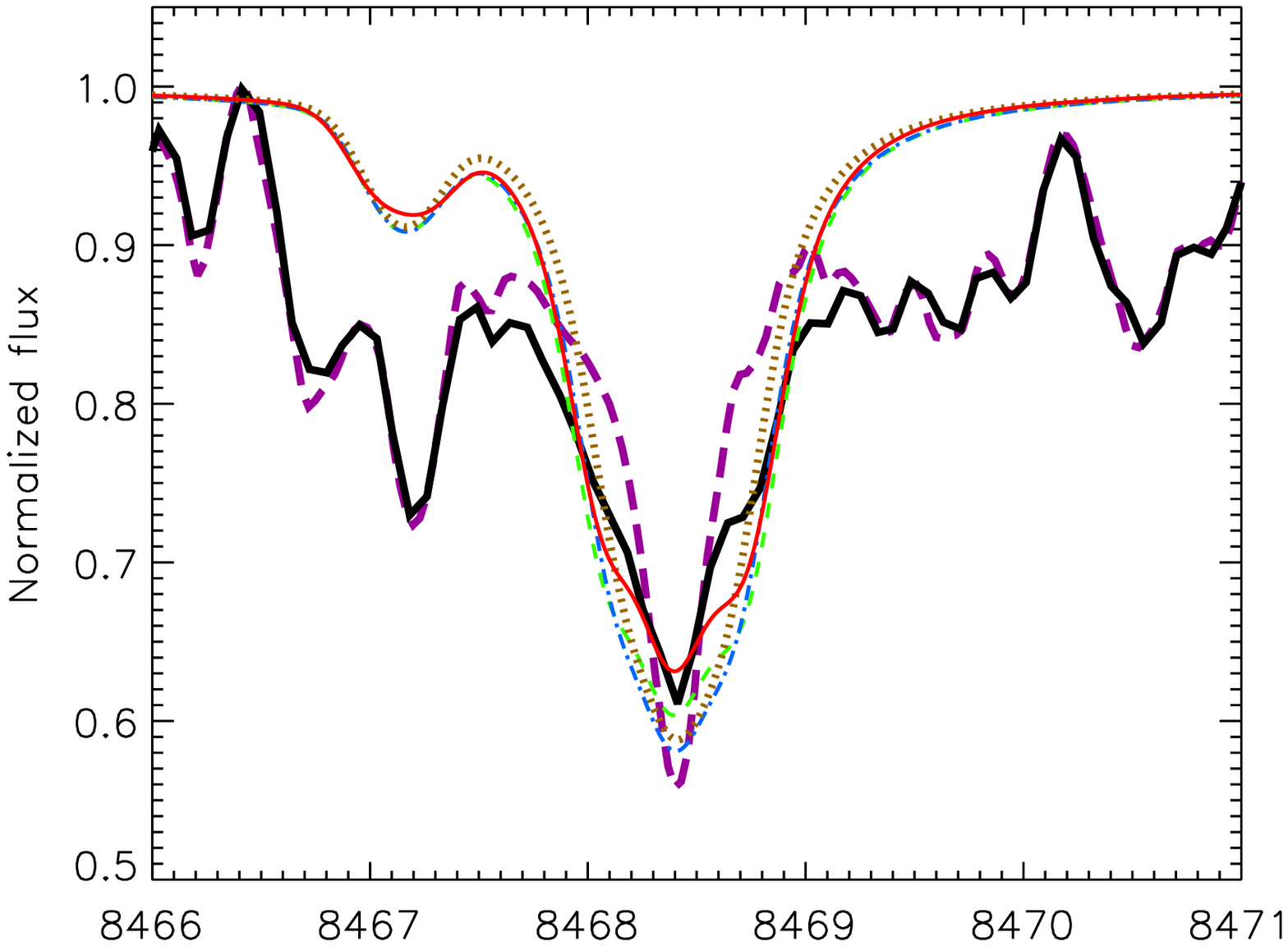}
\includegraphics[width=0.5\hsize]{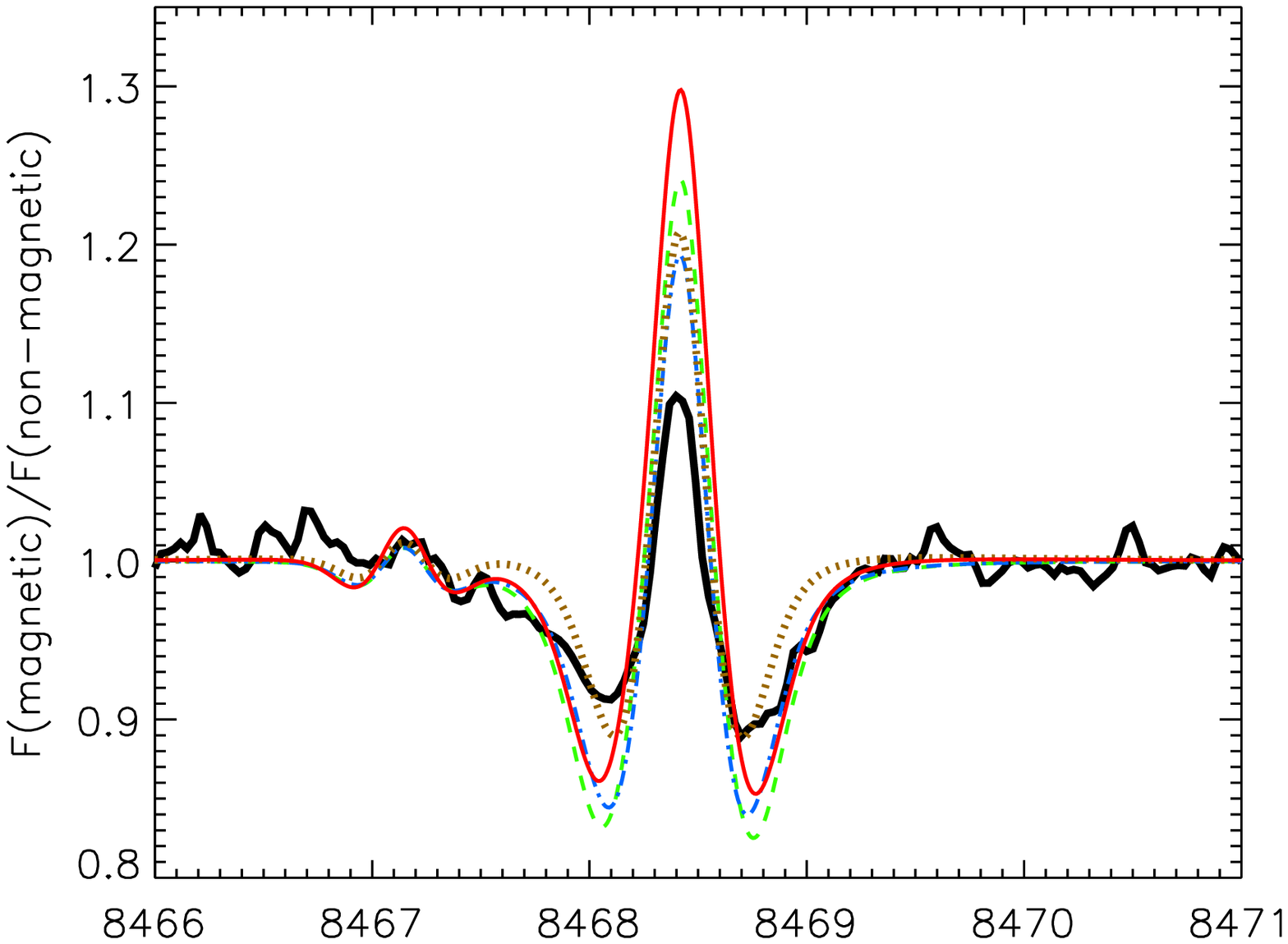}\\\\
\includegraphics[width=\hsize]{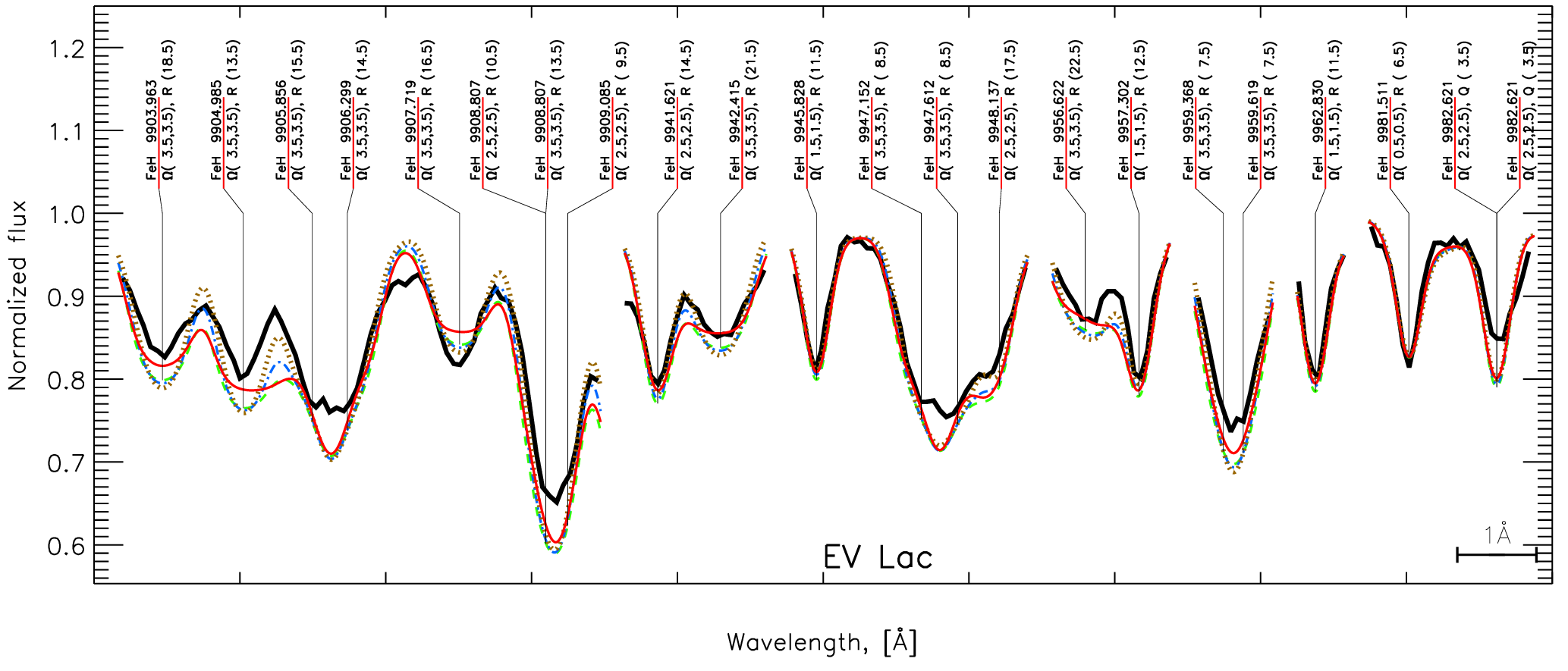}
\caption{Same as Fig.~\ref{fig:yzcmi}, but for the active M$4.5$ dwarf EV~Lac.
Model parameters: $\teff=3400$~K, $\logg=5.0$, $\varepsilon_{\rm Fe}=-4.40$, \vsini$=1$~\kms. Thick black line~--~observations,
violet long-dashed line (upper left plot only)~--~observations of inactive Gl~682,
green dashed~--~ $\bv=(0,4,0)$~kG, blue dash-dotted~--~$\bv=(2.5,2.5,0)$~kG, red solid~--~ $\bv=(4,0,0)$~kG,
dotted~--~$\bv=(3,0,0)$~kG. Wavelengths are in air for the upper plot and in vacuum for the lower.}
\label{fig:gl873}
\end{figure*}

\begin{figure*}
\includegraphics[width=0.5\hsize]{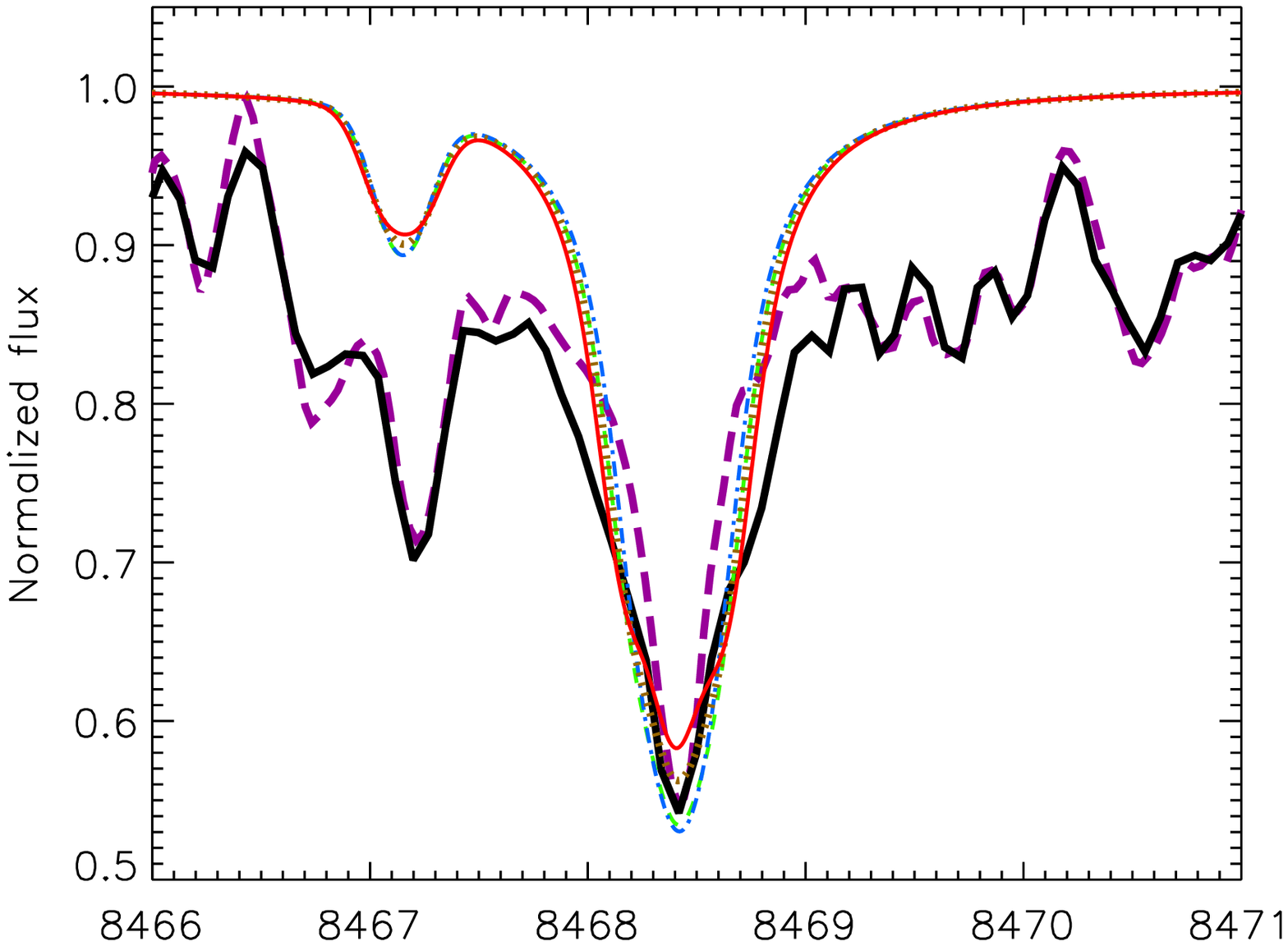}
\includegraphics[width=0.5\hsize]{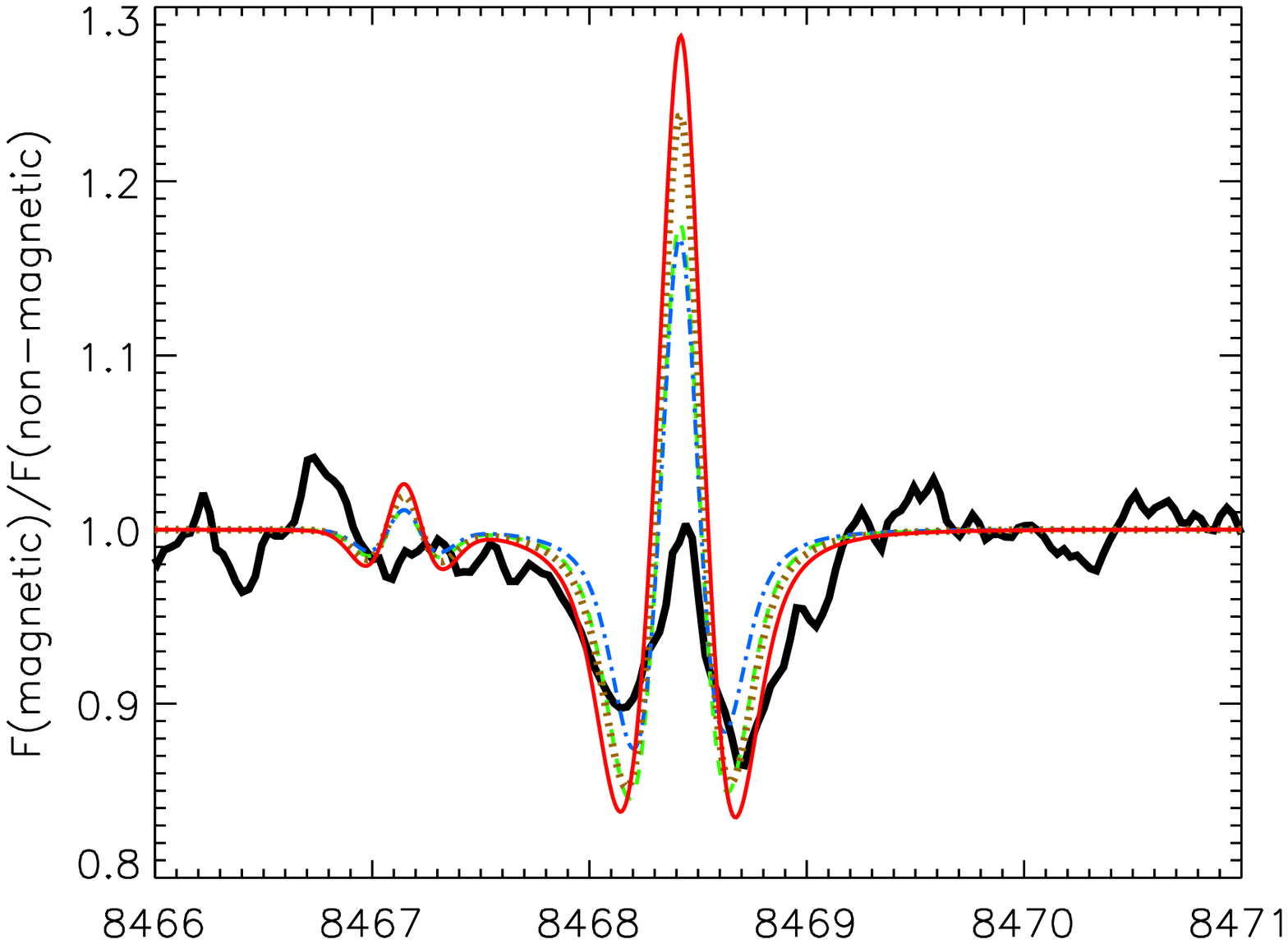}\\\\
\includegraphics[width=\hsize]{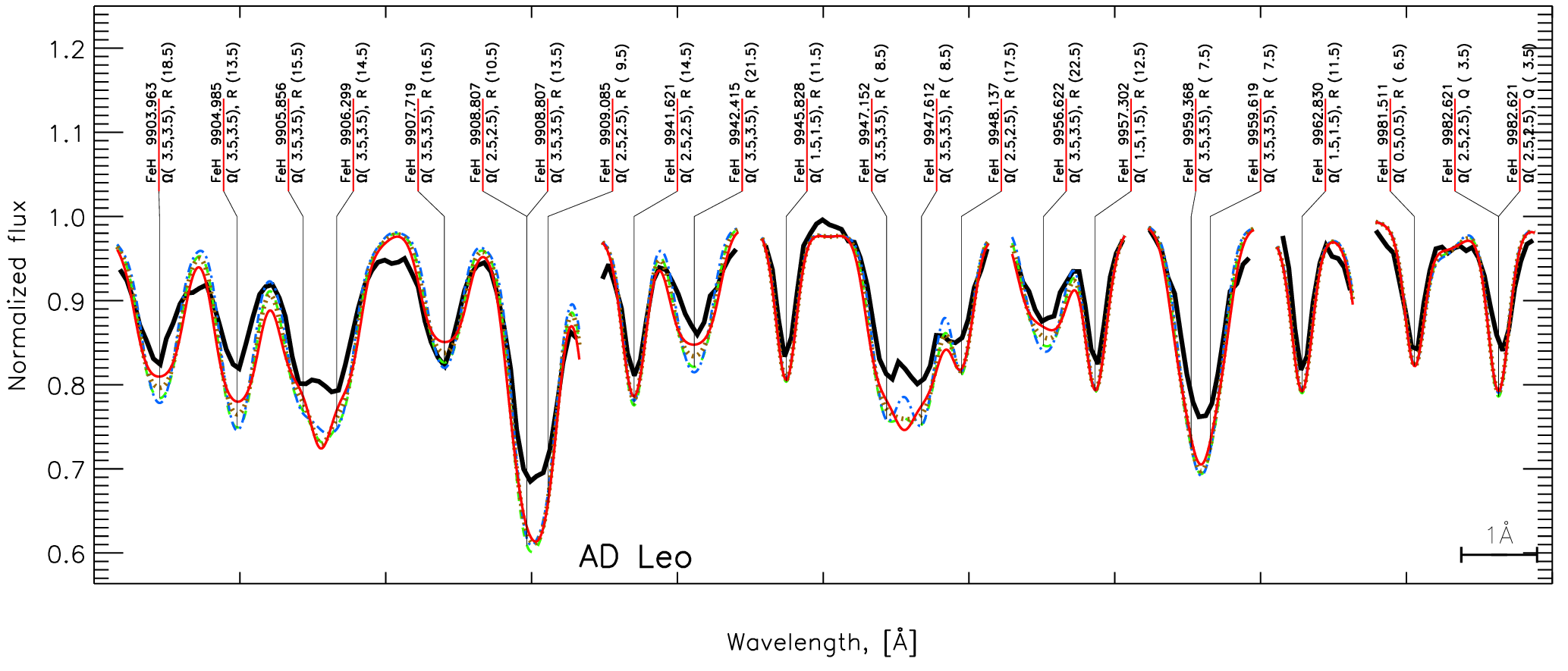}
\caption{Same as Fig.~\ref{fig:gl873}, but for the active M$3.5$ dwarf AD~Leo.
Model parameters: $\teff=3400$~K, $\logg=5.0$, $\varepsilon_{\rm Fe}=-4.50$, \vsini$=3$~\kms. Thick black line~--~observations,
violet long-dashed line (upper left plot only)~--~observations of inactive Gl~682,
green dashed~--~ $\bv=(1.7,1.7,0)$~kG, blue dash-dotted~--~$\bv=(2,0,0)$~kG, brawn dotted~--~$\bv=(2.5,0,0)$~kG,
red solid~--~ $\bv=(2.9,0,0)$~kG. 
Wavelengths are in air for the upper plot and in vacuum for the lower.}
\label{fig:adleo}
\end{figure*}

\subsubsection{EV~Lac}

The dwarf EV~Lac (GJ~873) is another object with a strong field that was previously reported \citep[see][]{jk-valenti2000,r-and-b-2007}. 
Again, the relatively low quality ($R=31\,000$) of the data makes it difficult to conclude on the field strength
from the analysis of the \ion{Fe}{i} $8468$~\AA\ line, as illustrated in Fig.~\ref{fig:gl873}, but at least its blue wing
is unaffected by instrumental effect.
As for YZ~CMi, a $\sim4$~kG magnetic field can still result in the position of a red-shifted $\sigma$-component 
in the line, as demonstrated by the theoretical calculation, but a field of $3.5$~kG seems to better match the position of 
the blue-shifted $\sigma$-component. Taking into account that a $3$~kG field provides probably the worst match, the actual
mean magnetic field is likely between $3$ and $4$~kG.
A sharp core of the Fe line itself (as a result of low-resolution observations) renders any quantitative 
measurement of the magnetic field redundant.
Again, similar to the case of YZ~CMi, the FeH lines (mostly those around $9906$~\AA) point to weaker fields of about $3-3.5$~kG. 
We are thus forced to conclude that spectra of a much better quality are highly required for more definite results.
The model parameters we used are: $\teff=3400$~K, $\logg=5.0$, [M/H]$=0.0$, \vsini$=1$~\kms, iron abundance $\varepsilon_{\rm Fe}=-4.40$.

\subsubsection{AD Leo}
The source AD~Leo (GJ~388) is the famous flare star and the last target in our sample. At first, \citet{jk-valenti2000} reported
a field of $\btimesf\sim3.3$~kG, which was then slightly decreased to $2.9$~kG in \citet{r-and-b-2007}.
Figure~\ref{fig:adleo} illustrates the fit to the FeH and \ion{Fe}{i} $8468$~\AA\ lines under the assumption
of different magnetic field intensity. The model parameters we used are: $\teff=3400$~K, $\logg=5.0$, $\varepsilon_{\rm Fe}=-4.50$, \vsini$=3$~\kms.
Note that it is impossible to match the width and depth of magnetic insensitive FeH lines with any
combination of \vsini, $\teff$, and Fe abundances 
unless one assumes a somewhat higher spectral resolution. We found
that $R\approx45\,000$ is required instead of $R=30\,000$. We adopt this higher
effective resolving power, which can be caused by the better image
quality at a seeing of the order of 0.8". This is likely to provide a
better effective resolution than the much wider slit size.
The profiles of the FeH lines suggest $\b=2-2.5$~kG on average, 
but with anomalously narrow line cores, whose fit would require an even lower field modulus
(again, with FeH~$9906$~\AA\ definitely pointing to $\b\sim2$~kG). It is hard to conclude precisely
about the field intensity from the \ion{Fe}{i} $8468$~\AA\ line, as seen from the upper panel
of Fig.~\ref{fig:adleo}.

\section{Discussion}
\label{sec:discussion}

In spite of recent progress in understanding and modelling the Zeeman effect in cool spectra of FeH
transitions, there are still problems which limit the precise measurements of the stellar magnetic fields.

The approximate treatment of the intermediate Hund's case is the main source of uncertainties in the analysis
of FeH lines.
It was noted already in \citet{afram2008} and confirmed in the present calculations that the effective
Hamiltonian of the intermediate case seems to miss an important term, which results in the underestimation of line broadening
as seen from theoretical profiles of magnetically sensitive lines.
It can only describe the Zeeman pattern of states with lower $\Omega$ numbers, and fails for the great majority of lines. 
This unknown perturbation still awaits its explanation. Nevertheless, we show that a semi-empirical approach of using different
Hund's cases for different states depending on their quantum numbers can very well reproduce profiles of many FeH
lines in the Wing-Ford band. This allows us to select lines with correctly reproduced Zeeman patterns and to use them
as probes of the magnetic fields in other stars by direct spectra synthesis.

In our semi-empirical approach we used a sunspot spectrum to
adjust theoretical g-factors of FeH lines. Because the properties of the plasma in a cool sunspot region are less accurately known
than the surrounding ``standard'' atmosphere of the Sun, this is where in principle some systematics could be introduced
in the resulting values of g-factors. However, these effects are minimized because a) the magnetic field inside
the spot and plasma properties are found from the independent diagnostics as provided by atomic lines,
b) the magnetic insensitive FeH lines are accurately fitted,
and c) we obtained a self-consistent fit to FeH and atomic lines using the same field strength, abundances, and model atmosphere.
In addition, we tried for the FeH lines to fit not only the line broadening but also the Zeeman pattern 
(i.e. shapes of theoretical and observed FeH lines).

Owing to the high $J$-numbers, it is impossible to see individual Zeeman components of molecular lines because they
are smeared out and result only in the line broadening (a single FeH line can give rise to up to few tens and even hundreds 
of $\pi$-  and $\sigma$-components). To accurately measure this broadening, high resolution and high $S/N$ 
(preferably $>100$) observations are desired. In this regard, and also because of our accurate knowledge of g-factors,
atomic lines are superior probes of magnetic fields with intensities higher than a few kG. For weaker fields, FeH lines
are a better choice, but, once again, under the condition that their splitting is well understood.

For two stars, YZ~CMi and EV~Lac, we confirm the presence of strong magnetic fields from the analysis of FeH spectra, 
but their intensity appears to be systematically lower than previously reported (see Fig.~\ref{fig:yzcmi}, for example). 
The forest of strong TiO features limits an accurate analysis of (ideally) good indicators such as 
\ion{Fe}{i} $8466$~\AA\ line. Furthermore, for both stars the low signal-to-noise ratio of the data forbids
an accurate fitting of the FeH lines. What remains are the widths of individual lines, which are very well fitted in the
sunspot spectra, providing in this way the more or less solid ground for the magnetic field analysis.
All these lines point to weaker fields, probably not more than $3.5$~kG. Even though a field of $4$~kG with a dominating
horizontal component gives line depths similar to those of $3.5$~kG, the line widths still appear to be wider than
observed. This is an expected result because the choice of the field geometry (with the same intensity)
does not affect the width of the line. This is another strong reason to call for more precise observations. 

The analysis of the FeH spectra of another active M$3.5$ star AD~Leo also points to a weaker (by $15-30$\%) mean magnetic field
than previously reported. 
For instance, from the analysis of an infrared \ion{Na}{i}~$2208.4$~\AA\ line \citet{kochukhov2009} 
find $\sum\btimesf=4.5$~kG and $\sum\btimesf=3.2$~kG for YZ~CMi and AD~Leo respectively. In the former case the field
is $27$\% larger than those derived in \citet{jk-valenti2000}, which can be a result of different probes
(\ion{Fe}{i}~$8468$~\AA\ and \ion{Na}{i}~$2208.4$~\AA) and techniques. 
This discrepancy between the magnetic fields inferred from the atomic and molecular diagnostics is unlikely 
to be caused by the different depths of line formation that might be modelled in model atmospheres with different accuracy. 
This is because a good fit to both atomic \ion{Ti}{i} and FeH lines is obtained for the non-magnetic
GJ~1002, which significantly reduces (but not necessary completely excludes!) any possible model atmosphere effect. 
More intense investigation is clearly needed
and should involve simultaneouse atomic and molecular probes for a stronger confirmation of our results.

Last but not least, the parameters of the FeH lines are derived from theoretical computations and thus may suffer from systematic inaccuracies. 
In this study we tried to adjust the parameters of some FeH with the spectra of non-magnetic stars, but this needs more
thorough investigation \citep[see][]{wende2010}.
Note again that the adjustment of the Einstein A-coefficients of some FeH lines from the original line list
has very little effect on the measured magnetic field intensities. This is because the latter relies on the
line broadening due to Zeeman splitting and not on the line depth, which is modified by corrected A-coefficients.
The inferred iron abundance is different in some cases from its recent solar value. This can be the result
of using precalculated model atmospheres with fixed effective temperatures: 
to fit the spectrum of a particular star we always used a model atmosphere
from the grid whose $\teff$(model) is close but can still be different from the $\teff$(star). In cool atmospheres
parameters such as $\teff$ and the Fe abundance are degenerate, i.e. the spectra of FeH react the same way 
if one changes any of these parameters. Thus, the solution is always not unique in this particular parameter's space.
Nevertheless, an uncertainty of about $0.2$~dex in Fe abundance (as in the case of our reference non-magnetic star GJ~1002), 
and any uncertainty in A-coefficients, 
would  mostly change the line depths, and therefore we expect them to have little effect
on the inferred magnetic field intensity, because this relies on the line broadening due to Zeeman splitting.

For GJ~1224, the splitting pattern of FeH lines points to the field intensity which is, at maximum, $1$~kG less 
than the one reported in \citet{r-and-b-2007}, which was $\btimesf=2.7$~kG. A $2.7$~kG field is probably too high as
demonstrated by FeH lines (see Fig.~\ref{fig:gj1224}). In particular, the separation of two 
FeH lines $9905$~\AA\ and $9906$~\AA\ appears to be a good indicator of the mean field intensity.
As noted above, with increasing field strength over $\approx2$~kG, these lines tend to produce a 
characteristic feature, which is not observed. Another interesting behaviour is demonstrated by the Ti lines red-ward of $1\,\mu$m.
The central depths of three of them ($10\,399.6$~\AA, $10\,498.9$~\AA, $10\,587.5$~\AA) require a field around $2.7$~kG, but
their widths, in contrast, point to a much weaker field. In addition, \ion{Ti}{i} $10\,610$~\AA, $10\,664.5$~\AA\ are 
well reproduced with $\approx2$~kG field.
That these lines are nicely fitted in the spectra of the non-magnetic GJ~1002 possibly excludes systematic
inaccuracies in spectra processing, because both spectra were observed with the same instrument and processed in the same way.

\begin{table}
\caption{Atmospheric parameters of investigated M-dwarfs.}
\label{tab:stellar}
\begin{footnotesize}
\begin{center}
\begin{tabular}{lcccrrr}
\hline
\hline
Name    & Spectral      & $\teff$ & \vsini & \multicolumn{2}{c}{$\b_{\mathrm{m}}$} & \multicolumn{1}{c}{$\btimesf$}\\
        &  type         & (K)     & (\kms) & \multicolumn{2}{c}{(kG)}              & \multicolumn{1}{c}{(kG)}\\
        &               &         &        & atoms             & FeH               &\\
\hline
Sunspot & --            & 4000    & 0.0    & $2.7$             & $2.7$             & $2.7^{(1)}$\\
GJ~1002 & M$5.5$        & 3100    & 2.5    & $0$               & $0$               & --\\
GJ~1224 & M$4.5$        & 3200    & 3.0    & $\approx2$        & $1.7-2$           & $2.7^{(2)}$\\
YZ~Cmi  & M$4.5$        & 3300    & 5.0    & $3-4$             & $3-3.5$           & $>3.9^{(2)}$\\
EV~Lac  & M$3.5$        & 3400    & 1.0    & $3-4$             & $3-3.5$           & $\sim3.9^{(3)}$\\
AD~Leo  & M$3.5$        & 3400    & 3.0    & $2-3$             & $2-2.5$           & $\sim2.9^{(2)}$\\
\hline
\end{tabular}
\end{center}
$\b_{\mathrm{m}}$~--~mean surface magnetic field\\
$\btimesf$~--~results of previous investigations\\\\
(1)~--~\citet{wallace1998}\\
(2)~--~\citet{r-and-b-2007}, scaled from (3)\\
(3)~--~\citet{jk-valenti1996}
\end{footnotesize}
\end{table}

There is, however, a consideration that relies on different dates of spectra taken for GJ~1224. 
Thus, if one assumes a presence of a magnetic spot(s) on the surface of the star and if the star is hosting
a strong field, and the rest of the surface has a somewhat weaker field, then a different field intensity observed in different
spectral regions and at different rotational phases is indeed naturally expected. The presence of wide regions with 
strong magnetic fields was confirmed for some M-dwarfs, but the field topology of the majority of them 
is found to be dominated by the poloidal component, which indicates rather homogeneous magnetic fields 
\citep[see][]{donati2006,donati2008,morin2008}. 
The complete characterization of the magnetic field in active M-dwarfs via the
phase-resolved observations in all Stokes parameters are thus of particular interest, 
though it is a very challenging task even for modern instruments.

In this work we attempted to measure a mean surface magnetic field. This may not correspond to a real picture because, as
stated above, the magnetic field of M-dwarfs is likely to be concentrated in magnetically active areas (spots) similar to what
is observed in the solar atmosphere. Thus, a mean surface field can be represented by (at least) two components: a strong field
inside a spot/spots and a much weaker component of a surrounding surface (which can be zero as well). To account for this geometrical
consideration the corresponding filling factors $f$ (i.e. free parameters that describe the relative size of magnetic regions characterized
by a given magnetic field modulus)
must be applied. The more accurate way to account for the complex magnetic field geometry without the need of filling factors
would be the analysis of polarized radiation in individual lines. However, as stated above, this is extremely difficult or
almost impossible to do with the current instrumentation.
Again, similar to the Sun, the temperature inside these spot(s)
may also be lower than the surrounding plasma. This will be investigated in our future works, also with data of better quality.

Table~\ref{tab:stellar} gathers the main results of the present study. An average surface magnetic field resulting from the analysis
of atomic and FeH lines are shown separately. A large scatter resulting from the analysis of the \ion{Fe}{i}~$8468$~\AA\ line
is due to the uncertainties of fitting blue and red wings of the flux ratio (see, for instance, Figs.~\ref{fig:gj1224} and \ref{fig:gl873}), 
which seem to require different $\b$.

\section{Summary}
\label{sec:summary}

In this study we made an attempt of using up-to-date knowledge of molecular Zeeman effect,
modern software for the magnetic spectra synthesis, and molecular lines data to develop an approach
of modelling the Zeeman splitting in FeH lines of Wing-Ford $F^4\,\Delta-X^4\,\Delta$ band.
This approach was then applied to measure magnetic fields in selected M-dwarfs
for which observations in both atomic and FeH lines are available. The main results of the present work
can be summarized as follows:
\begin{enumerate}
\item
Our results of the magnetic field strengths derived from FeH lines are $15-30$\%\ lower than results presented in
\citet{r-and-b-2007}, which are based on atomic line analysis scaled from \citet{jk-valenti2000}.
\item
We confirm a strong magnetic field found in YZ~CMi, EV~Lac, and AD~Leo but with an
intensity which is likely $500$~G (or more) weaker. Unfortunately, the poor quality of the data forbids a more quantitative conclusions.
\item
An analysis of atomic and FeH lines in spectra of GJ~1224 points in the direction of $1.7-2$~kG averaged
magnetic field, which is lower than $2.7$~kG previously reported \citep{r-and-b-2007}.
\item
The estimates of the magnetic field modulus from the \ion{Fe}{i}~$8468$~\AA\ line seem to be systematically higher 
than those from FeH lines. This, however, should be taken with caution because of unknowns associated with
the quality of the data and atmospheric parameters used.
\item
With routines provided by Molecular Zeeman Library we developed an algorithm for calculating
Land\'e g-factors of FeH states: the choice of the Hund's cases (a), (b) or intermediate follows
from the consideration of quantum numbers of individual states. The test computations of the sunspot
spectra showed a good agreement with observations and with calculations based on the best-fit g-factors
from \citet{afram2008}.
\item
To distinguish between different magnetic field geometries, higher quality observations are required 
for the analysis of FeH spectra ($S/N>100$).
\item
To exclude possible selection effects caused by the variable field associated with the magnetic spot(s), and to provide
a complete characterization of the magnetic field intensity and geometry, time-resolved observations are highly desired.
These observations, which would ideally provide spectra in all four Stokes parameters as well, 
remain highly difficult even for the brightest M-dwarfs because they require 
many hours of signal integration time to achieve the desired S/N even with modern  astronomical instrumentation.
\end{enumerate}

\begin{acknowledgements}
We thank Dr. Bernard Leroy for his kind advice and help with the MZL routines. 

This work was supported by the following grants: Deutsche Forschungsgemeinschaft (DFG)
Research Grant RE1664/7-1 to DS and Deutsche Forschungsgemeinschaft under DFG RE 1664/4-1 and NSF
grant AST07-08074 to AS. SW acknowledges financial support from the DFG
Research Training Group GrK - 1351 "Extrasolar Planets and their host stars".
OK is a Royal Swedish Academy of Sciences Research Fellow supported 
by grants from the Knut and Alice Wallenberg Foundation and the Swedish Research Council.
We also acknowledge the use of electronic databases (VALD, SIMBAD, NASA's ADS).
This research has made use of the Molecular Zeeman Library (Leroy, 2004).
\end{acknowledgements}

%
\listofobjects
\end{document}